\documentclass[12pt]{article}

\usepackage[utf8]{inputenc} 

\usepackage{geometry} 
\usepackage{graphicx} 

\usepackage{booktabs} 
\usepackage{array} 
\usepackage{paralist} 
\usepackage{verbatim} 
\usepackage{amsbsy}
\usepackage{amsmath}
\usepackage{longtable}
\usepackage[toc,page]{appendix}
\usepackage{natbib}
\usepackage{url}
\RequirePackage[colorlinks,citecolor=blue,urlcolor=blue]{hyperref}
\usepackage{breakurl}

\usepackage{amssymb}

\usepackage[textsize=small]{todonotes}
\usepackage{subcaption}

\usepackage{multirow}

\usepackage{dcolumn}

\raggedright
\graphicspath{{figures/}}

\renewcommand{\tableofcontents}{}

\bibpunct{(}{)}{;}{a}{}{,}  

\begin{document}

\bigskip
\medskip
\begin{center}

\noindent{\Large \bf Markov-modulated continuous-time Markov chains} \\
\noindent{\Large \bf to identify site- and branch-specific} \\
\noindent{\Large \bf evolutionary variation}

\bigskip

\noindent{\normalsize \sc
	Guy Baele$^{1}$,
	Mandev S. Gill$^{1}$,
	Philippe Lemey$^{1}$, \\
    and Marc A.~Suchard$^{2,3,4}$}\\
\noindent {\small
  \it $^1$Department of Microbiology, Immunology and Transplantation, Rega Institute, KU Leuven, Herestraat 49, 3000 Leuven, Belgium \\
  \it $^2$Department of Biostatistics, Jonathan and Karin Fielding School of Public Health, University
  of California, Los Angeles, United States} \\
  \it $^3$Department of Biomathematics, David Geffen School of Medicine at UCLA, University of California,
  Los Angeles, United States \\
  \it $^4$Department of Human Genetics, David Geffen School of Medicine at UCLA, Universtiy of California,
  Los Angeles, United States

\end{center}
\medskip
\noindent{\bf Corresponding author:} Guy Baele, Department of Microbiology, Immunology and Transplantation, Rega Institute, KU Leuven, Herestraat 49, 3000 Leuven, Belgium; E-mail: \url{guy.baele@kuleuven.be}\\



\newcommand{\psaB}{\emph{psaB }}
\newcommand{\ndhD}{\emph{ndhD }}

\newcommand{\numTaxa}{N}
\newcommand{\numColumns}{C}
\newcommand{\columnIdx}{c}
\newcommand{\branchIdx}{b}
\newcommand{\branchLength}[1]{t_{#1}}
\newcommand{\ctmcProbability}[3]{\mathbb{P}_{#1 #2}( #3 )}
\newcommand{\asrvMultiplier}[1]{r_{#1}}
\newcommand{\branchMultiplier}[1]{\mu_{#1}}

\newcommand{\order}[1]{{\cal O} \left( #1 \right)}

\newcommand{\numStates}{S}
\newcommand{\stateIdx}{s}
\newcommand{\stateIdxTwo}{\stateIdx^{\prime}}

\newcommand{\numModels}{K}
\newcommand{\modelIdx}{k}
\newcommand{\modelIdxTwo}{\modelIdx^{\prime}}

\newcommand{\mmRateMatrix}{\mathbf{\Lambda}}
\newcommand{\rateMatrix}[1]{\mathbf{Q}_{#1}}
\newcommand{\probMatrix}[1]{\mathbf{P}_{#1}}
\newcommand{\rateMatrixElement}[3]{Q^{(#1)}_{#2 #3}}
\newcommand{\stationaryDistribution}[2]{\pi_{#1 #2}}

\newcommand{\switchingRate}[2]{\phi_{#1 #2}}
\newcommand{\identityMatrix}[1]{\mathbf{I}_{#1}}
\newcommand{\switchingMatrix}[2]{\switchingRate{#1}{#2} \identityMatrix{}}
\newcommand{\rateMultiplier}[1]{\rho_{#1}}
\newcommand{\scaledRateMatrix}[1]{\rateMultiplier{#1} \rateMatrix{#1}}
\newcommand{\tm}{\text{-}}
\newcommand{\regimeMatrix}{\mathbf{\Phi}}

\newcommand{\diagonalRateMatrix}[1]{\coloredBox{\scaledRateMatrix{#1}}
- \sum\limits_{\modelIdx \neq #1} \switchingMatrix{#1}{\modelIdx}
}

\newcommand{\bigzero}[2]{\makebox(0,0){\hspace{#1}\vspace{#2}\text{\huge0}}}

\newcommand{\coloredBox}[1]{%
\raisebox{-0.5em}{
\tikz \node[
		fill=orange!20,
		align=center,
		minimum width={%
		3em
		}
	] (0,0) {%
\ensuremath{#1}\hspace{-0.5em} %
}; %
} 
}

\paragraph{Abstract}

Markov models of character substitution on phylogenies form the foundation of phylogenetic inference frameworks. Early models made the simplifying assumption that the substitution process is homogeneous over time and across sites in the molecular sequence alignment.
While standard practice adopts extensions that accommodate heterogeneity of substitution rates across sites, heterogeneity in the process over time in a site-specific manner remains frequently overlooked.
This is problematic, as evolutionary processes that act at the molecular level are highly variable, subjecting different sites to different selective constraints over time, impacting their substitution behaviour.
We propose incorporating time variability through Markov-modulated models (MMMs) that allow the substitution process (including relative character exchange rates as well as the overall substitution rate) that models the evolution at an individual site to vary across lineages.
We implement a general MMM framework in BEAST, a popular Bayesian phylogenetic inference software package, allowing researchers to compose a wide range of MMMs through flexible XML specification.
Using examples from bacterial, viral and plastid genome evolution, we show that MMMs impact phylogenetic tree estimation and can substantially improve model fit compared to standard substitution models.
Through simulations, we show that marginal likelihood estimation accurately identifies the generative model and does not systematically prefer the more parameter-rich MMMs.
In order to mitigate the increased computational demands associated with MMMs, our implementation exploits recently developed updates to BEAGLE, a high-performance computational library for phylogenetic inference.

\clearpage

\section{Introduction}

Molecular sequence evolution is typically modeled by Markov models of character substitution acting along the branches of a phylogenetic tree.
These models are phenomenological descriptions of the evolution of DNA as a string of a number of discrete character states, with models of nucleotide substitution among four states being the most widely used in statistical phylogenetics.
The Markovian property within such a model reflects the common assumption that evolution has no memory.
Further, it is standard to assume that  the Markov model is time-homogeneous, so that it can be characterized by a generator or instantaneous rate matrix $\rateMatrix{}$ that remains constant during evolution \citep{gascuel2007modelling}.
Starting with the original nucleotide substitution model of Jukes and Cantor \citep{Jukes1969}, which does not require any parameter to be estimated from the data, a number of different Markov models of DNA sequence evolution have been proposed over the years, of which the HKY model \citep{HKY} and the general time-reversible model \citep{Lanave1984,Tavare1986} are among the most widely used.

The stochastic process $\{X(t) : t \geq 0\}$ over discrete state space $\mathcal{S}$ is called a continuous-time Markov chain (CTMC) if for all $t \geq 0, b \geq 0$ and $ i,j \in \mathcal{S}$,
\begin{equation}
\label{eq:markov}
P(X(t+b) = j \mid X(b) = i, \{ X(u) : 0 \leq u < b \}) = P(X(t+b) = j \mid X(b) = i).
\end{equation}
In other words, a CTMC is a stochastic process having the Markovian property that the conditional distribution of the future $X(t+b)$ given the present $X(b)$ and the past $X(u), 0 \leq u < b$, depends only on the present and is independent of the past \citep{Tolver2016}.
If additionally, $P(X(t+b) = j \mid X(b) = i)$ is independent of $b$, then the CTMC has stationary or time-homogeneous transition probabilities.
In this case, we have
\begin{equation}
\label{eq:markovhomog}
P(X(t+b) = j \mid X(b) = i) = P(X(t) = j \mid X(0) = i),
\end{equation}
and we denote this transition probability by $P_{ij}(t)$.
The transition probabilities for all $i,j \in \mathcal{S}$ can be obtained through computing the matrix exponential
\begin{equation}
\label{eq:matrixexp}
\probMatrix{}(t) = \mathrm{exp}(\rateMatrix{}t),
\end{equation}
which is generally accomplished via diagonalization of $\rateMatrix{}$ \citep{Felsenstein2004} and leads to an $S \times S$ matrix called the transition probability matrix where $S = | \mathcal{S} |$.

Early probabilistic phylogenetic reconstruction methods assumed a single substitution model that acted independently across all sites and lineages.
However, the evolutionary processes that act at the molecular level are highly variable, as some sequence positions evolve rapidly while others barely undergo any changes.
Different sites are subject to different selective constraints, as exemplified for different codon positions in protein coding genes where third codon positions usually evolve faster than first positions, which, in turn, evolve faster than second positions.
Accommodating different rates at different sites has been one of the most influential modeling innovations in molecular phylogenetics \citep{Yang1996}. 
However, sites may also evolve in a qualitatively different manner, in which case simply accommodating among-site rate variation is insufficient.
\citet{Pagel2004} incorporated such ``pattern heterogeneity'' through a flexible mixture model that posits that, among a collection of prespecified substitution models, each substitution model applies to any given alignment site with a specific probability, or ``weight.''
The likelihood for each site is then the weighted sum over these different substitution models.
Notably, the \citet{Pagel2004} mixture model for pattern heterogeneity can be combined with the popular among-site rate variation model of \citet{yang1994maximum} that assumes site-specific substitution rates drawn from a discretized gamma distribution.


Markov-modulated models (MMMs) naturally extend mixture models to account for time variability, and were introduced into phylogenetics \citep{tuffley1998modeling,Lockhart1998} to accommodate observations that the evolutionary rate of a particular site in coding sequences can vary across the phylogeny because sites that are critical for the function of a macromolecule may change within the nucleotide sequence over time \citep{Fitch1970}.
MMMs allow the process that governs the evolution of an individual site to change over time along a phylogeny by augmenting the usual state space of the substitution model (consisting of sequence characters such as nucleotides or codons) to include categories corresponding to different evolutionary models.
These categories also change following a homogeneous, stationary and time-reversible Markovian process, similar to standard CTMCs of character substitution.
\citet{gascuel2007modelling} presented a general time-reversible model (GTR) for $K$ categories:
\begin{equation}
\label{eq:switchgtr}
\regimeMatrix_{GTR} = \delta \left(
				\begin{array}{ccccc}
					-          & \eta_{2}R_{1 \leftrightarrow 2}            & \cdots                                        & \eta_{K}R_{1 \leftrightarrow K}              \\
					\eta_{1}R_{1 \leftrightarrow 2}          & - &                                               & \vdots                                        \\
					\vdots                          &                                   & \ddots                                        & \eta_{K - 1}R_{K -1 \leftrightarrow K} \\
					\eta_{1}R_{1 \leftrightarrow K} & \cdots                            & \eta_{K -1}R_{K -1 \leftrightarrow K} & -
				\end{array}
				\right),
\end{equation}
\newcommand{\categoryDistribution}{\psi}
where each row sums to 0, $\boldsymbol{\eta} = \left(\eta_1, \ldots, \eta_K\right)$ is the stationary distribution of the categories and $\delta$ is an additional parameter that expresses the global rate of changes between all categories.
Typically, the $R$ coefficients are normalized such that $\delta$ is the expected number of category changes during one time unit \citep{gascuel2007modelling}.
From Equation \ref{eq:switchgtr}, it follows that an MMM reduces to a mixture model with weights $\boldsymbol{\eta}$ when the rate $\delta$ at which changes between categories occur tends to zero \citep{guindon2004modeling,gascuel2007modelling}.

Nearly fifty years ago, \citet{Fitch1970} proposed their ``concomitantly variable codons'' (or ``covarion'') hypothesis, which states that at any given time some sites are invariable due to functional or structural constraints, but that as mutations are fixed elsewhere in the sequence these constraints may change, so that sites that were previously invariable may become variable and vice versa.
The covarion hypothesis therefore implies that the pool of variable sites is changing with time.
Inspired by these observations, 
\citet{tuffley1998modeling} put forth a simple covarion-style model where a two-state Markov process acts as a switch that turns sites ``on'' (variable) and ``off'' (invariable), but does not trigger a concomitant reaction in any other sites.
Under their model, substitutions occur at a fixed rate for sites that are turned ``on'', while no substitutions occur for sites that are turned ``off.''
\citet{Huelsenbeck2002} generalized this model by allowing each site to have a fixed rate drawn from a discrete gamma distribution, with sites assuming the fixed rate when turned ``on'' and assuming a zero rate when turned ``off''.
\citet{Zhou2010} further extended the \citet{Huelsenbeck2002} model by relaxing the assumption that the two category transition rates (one from ``on'' to ``off'', and the other from ``off'' to ``on'') are constant across sites, employing a Dirichlet process mixture to achieve rate heterogeneity.
\citet{galtier2001maximum} presented an alternative generalization of the \citet{tuffley1998modeling} model by assuming that a fixed subset of sites evolves according to a covarion process, with rates drawn from a discrete gamma distribution and switching allowed between different rate classes at a single fixed rate.
The remaining sites that do not evolve under a covarion process are posited to have site-specific rates also drawn from a discrete gamma distribution.
\citet{Wang2007,Wang2009} introduced a general covarion model that combines the features of the \citet{tuffley1998modeling}, \citet{galtier2001maximum} and \citet{Huelsenbeck2002} models.

These covarion-style models can capture variations in lineage-specific evolutionary rates over time and as such constitute special instances of the Markov-modulated class of Markov processes \citep{fischer1993markov}, whose generator matrices are well described by Kronecker product formalism \citep{galtier2004markov}.
These models suffer from potentially high dimensionality, as the transition matrix has a dimension equal to the product of the number of character states and the number of rate categories, making matrix exponentiation and likelihood calculation computationally challenging.
To address these difficulties, \citet{galtier2004markov} proposed a fast algorithm for diagonalizing the generator matrix.
The authors show their algorithm to be tractable even for a large number of rate classes and a large number of states, employing up to 20 rate classes combined with full $61 \times 61$ codon models, leading to a $1220 \times 1220$ generator matrix of their MMM process.

The development of covarion-style MMMs has inspired the introduction of more general MMMs that permit switching between evolutionary models that differ in relative character exchange rates as well in overall substitution rate.
\citet{guindon2004modeling} introduced a codon-based MMM that allows substitutions between codons to occur under either a purifying, neutral, or positive selection regime and allowing changes (or switches) between selection classes to occur according to a three-state continuous-time Markov process.
The authors 
showed in an analysis of eight HIV-1 \emph{env} sequence data sets that their model provides a significantly better fit to the data than a model that does not accommodate switches between selection patterns.
\citet{gascuel2007modelling} proposed an extension to a general MMM framework for the analysis of nucleotide, amino acid or codon data.
Notably, they allow for switching between any evolutionary models, so long as all models share the same stationary distribution of characters.
\citet{Whelan2008} proposed a similar general MMM framework for nucleotide substitution that also permits the different substitution models to have different stationary distributions.

In spite of all of these developments, MMMs remain relatively under-utilised due to their prohibitive computational cost and a lack of readily accessible implementations in widely used software packages for phylogenetic inference.
There may also exist a general unfamiliarity with how MMMs may improve phylogenetic tree reconstruction.
We introduce a Bayesian inference framework for MMMs, implemented in BEAST \citep{beastX}, a software package for Bayesian evolutionary analysis.
We strive for optimal generality by allowing switching between evolutionary models that have different substitution rates, relative character exchange rates and stationary distributions.
Further, our implementation accommodates phylogenetic uncertainty and is equipped for analysis of any data type. 
Researchers can easily compose and customize MMMs through flexible XML specification, as well as use an MMM as part of a larger Bayesian phylodynamic analysis by combining it with any of the wide range of models for demographic reconstruction, phylogeographic inference, and phenotypic trait evolution available in BEAST.
To cope with the higher computational burden imposed by MMMs, our implementation leverages the power of the BEAGLE library for high-performance likelihood computation \citep{beagle3}, and we show considerable performance increases when employing graphics cards aimed at scientific computing.
We illustrate the flexibility and utility of our MMM implementation by employing a variety of different model compositions in examples of different complexity from bacterial, plastid, and viral evolution.
To the best of our knowledge, no comparably general MMM framework is currently available in a phylogenetic inference software package.

\section{Methods}

\subsection{Markov-modulated model structure}

Consider an MMM composed of $\numModels$ evolutionary models (irrespective of those models being nucleotide, amino acid or codon models).
Each evolutionary model is defined by a relative substitution rate multiplier $\rateMultiplier{\modelIdx}$ and a substitution model characterized by an instantaneous rate matrix $\rateMatrix{\modelIdx} = \left\{
			\rateMatrixElement{\modelIdx}{\stateIdx}{\stateIdxTwo}
			\right\}$, of dimension $\numStates$, and stationary distribution
			$\boldsymbol{\Pi}_{\modelIdx} = \left(
			\stationaryDistribution{\modelIdx}{1}, \ldots, \stationaryDistribution{\modelIdx}{\numStates}
			\right)$.
			We also adopt the usual constraint
			-$\sum_{\stateIdx = 1}^{\numStates} \rateMatrixElement{\modelIdx}{\stateIdx}{\stateIdx}
			\stationaryDistribution{\modelIdx}{\stateIdx} = 1$.
The switching process between the $\numModels$ models is defined by a $\numModels$-state continuous-time Markov process with rate matrix
\begin{equation}
\label{eq:switching}
\regimeMatrix = \left(
				\begin{array}{ccccc}
					-\sum_{k\neq1}\phi_{1k}          & \phi_{12}            & \cdots                                        & \phi_{1\numModels}              \\
					\phi_{21}          & -\sum_{k\neq2}\phi_{2k} &                                               & \vdots                                        \\
					\vdots                          &                                   & \ddots                                        & \phi_{\numModels -1,\numModels} \\
					\phi_{\numModels 1} & \cdots                            & \phi_{\numModels, \numModels -1} & -\sum_{k\neq\numModels}\phi_{\numModels k}
				\end{array}
				\right),
\end{equation}
where the element $\phi_{ij}$ corresponds to the rate of switching from substitution model $i$ to substitution model $j$, and the diagonal elements are fixed such that the rows sum to $0$.
We denote the stationary distribution of this switching process by $\boldsymbol{\Psi} = \left(\categoryDistribution_1, \ldots, \categoryDistribution_K\right)$. 
These model switches follow a homogeneous, stationary - but not necessarily time-reversible - Markovian process.
In Equation \ref{eq:switching}, we do not make use of an additional parameter $\delta$ that expresses the global rate of change between the evolutionary models (as in equation \ref{eq:switchgtr}) because this is a deterministic parameter obtained by normalizing the model-switching process \citep{guindon2004modeling,gascuel2007modelling}.

The MMM is characterized by a $\numModels\numStates \times \numModels\numStates$ rate matrix $\mmRateMatrix $\citep{fischer1993markov}:
\begin{equation}
\label{eq:singlematrix}
	\begin{aligned}
			\mmRateMatrix &=
				\left(
				\begin{array}{ccccc}
					\diagonalRateMatrix{1}          & \switchingMatrix{1}{2}            & \cdots                                        & \switchingMatrix{1}{\numModels}               \\
					\switchingMatrix{2}{1}          & \diagonalRateMatrix{2} &                                               & \vdots                                        \\
					\vdots                          &                                   & \ddots                                        & \switchingMatrix{\numModels\tm1,}{\numModels} \\
					\switchingMatrix{\numModels}{1} & \cdots                            & \switchingMatrix{\numModels,}{\numModels\tm1} & \diagonalRateMatrix{\numModels}
				\end{array}
				\right)
				\\
			&= \text{diag}\left(\scaledRateMatrix{1}, \ldots, \scaledRateMatrix{\numModels} \right)
			+ \regimeMatrix \otimes \identityMatrix{}, 
		\end{aligned}
\end{equation}
where $\identityMatrix{}$ is an $\numStates \times \numStates$ identity matrix and $\otimes$ denotes the Kronecker product.
The MMM can therefore be considered a single Markov process with a state space equal to the Cartesian product of the state space of the switching process (the evolutionary models) and the state space of the evolutionary models, with cardinality $\numModels \numStates$ and stationary distribution $\boldsymbol{\Pi}_{\mmRateMatrix} = \left( \categoryDistribution_{1}\pi_{11}, \ldots, \categoryDistribution_{1}\pi_{1\numStates}, \ldots, \categoryDistribution_{\numModels}\pi_{\numModels1}, \ldots, \categoryDistribution_{\numModels}\pi_{\numModels\numStates} \right)$ \citep{guindon2004modeling}.
As noted by \citet{gascuel2007modelling}, the MMM in Equation \ref{eq:singlematrix} allows for every compound state $(\modelIdx, \stateIdx)$ to either: (1) stay in model $\modelIdx$ and transition to $(\modelIdx, \stateIdxTwo)$ with rate defined by $\scaledRateMatrix{\modelIdx}$, or (2) change evolutionary models and transition to $(\modelIdxTwo, \stateIdx)$ with rate $\switchingRate{\modelIdx}{\modelIdxTwo}$.
All rows in $\mmRateMatrix$ sum to $0$, and because $\boldsymbol{\Psi}\regimeMatrix = 0$ and $\boldsymbol{\Pi}_{\modelIdx}\rateMatrix{\modelIdx} = 0$, it follows that $\boldsymbol{\Pi}_{\mmRateMatrix}\mmRateMatrix = 0$.


Standard nucleotide and codon substitution models are often superimposed with a model that employs a discretized gamma distribution to accommodate among-site rate variation \citep{yang1994maximum}. 
We will refer to this specific model as the ASRV model.
Combining the ASRV model with MMMs can result in a model that is not identifiable.
For example, consider a rooted tree of two taxa, $\numModels = 2$ with $\rateMatrix{1} = \rateMatrix{2}$ and an ASRV model with two rate categories with rate multipliers $\rateMultiplier{1}$ and $\rateMultiplier{2}$.
Further, assume a switching process with instantaneous rates $\switchingRate{1}{2} = \switchingRate{2}{1} = \switchingRate{}{}$ and hence equilibrium frequencies $\categoryDistribution_{1} = \categoryDistribution_{2} = 1/2$.
Given this setup, it becomes impossible to distinguish between sites that evolve according to model $\rateMatrix{1}$ (i.e.~when the switching process puts them in the first category) with rate $\rateMultiplier{2}$ and sites that evolve according to model $\rateMatrix{2}$ with rate $\rateMultiplier{1}$.

However, we can still superimpose the ASRV model on an MMM while maintaining identifiability by restricting the MMM relative rate multipliers:
\begin{equation}
	\rateMultiplier{1} = \cdots = \rateMultiplier{\numModels} = 1.
\end{equation}
The resulting model is a generalization of the model proposed by \citet{guindon2004modeling}, but with a more general switching process.
Note that the covarion model of \citet{tuffley1998modeling} is also nested within the general MMM formulation by setting $\numModels = 2$, $\rateMultiplier{1} = 0$ and $\rateMultiplier{2} = 1$.
The covarion model proposed by \citet{galtier2004markov} is also nested within our framework by positing one possible substitution model, i.e.
\begin{equation}
	\rateMatrix{\modelIdx} = \rateMatrix{} \text{ for all } \modelIdx.
\end{equation}

We note that our proposed MMM can be parameterized as the ubiquitously-used ASRV model, which employs a discretized gamma distribution to determine its relative rates \citep{yang1994maximum}, solely through setting its parameters as follows:
\begin{equation}
	\begin{aligned}
		\rateMatrix{\modelIdx} &= \rateMatrix{} \text{ for all } \modelIdx, \\
		\switchingRate{\modelIdx}{\modelIdxTwo} &= 0 \text{ for all } \modelIdx, \modelIdxTwo
	\end{aligned}
\end{equation}
and prescribing that $\rateMultiplier{\modelIdx}$ for all $\modelIdx$ equal the $(\modelIdx - \frac{1}{2}) / \numModels \times 100^{\text{\tiny\,th}}$ percentile of a gamma distribution with fixed expectation 1 and unknown shape $\alpha$.

Other common restrictions on the switching process, as extensively used in telecommunication and teletraffic networks \citep{fischer1993markov}, include an ordered model random walk:
\begin{equation}
	\switchingRate{\modelIdx}{\modelIdxTwo} = 0 \text{ for all } |\modelIdx - \modelIdxTwo| > 1,
\end{equation}
and an ordered model random walk with single rate:
\begin{equation}
	 		\switchingRate{\modelIdx}{\modelIdxTwo} =
	 			\left\{
	 			\begin{array}{l}
	 			\switchingRate{}{} \text{ if } |\modelIdx - \modelIdxTwo| = 1, \\
	 			0 \text{ otherwise. }
	 			\end{array}
	 			\right.
\end{equation}

\subsection{Likelihood}

In this section, we adopt a similar notation to \citet{gascuel2007modelling} to describe the data likelihood under an MMM.
Likelihood calculations for MMMs employ a standard pruning approach \citep{Felsenstein1981}, with integration over the compound states (i.e.~the evolutionary model and character state) 
at the internal nodes of the tree, and integration over the unobserved categories at the tips.
Let $\mathbf{Y} = (\mathbf{Y}_1, \ldots, \mathbf{Y}_{L})$ where $\mathbf{Y}_{\ell}$ are the extant characters observed at aligned site $\ell$ for $\ell = 1, \ldots, L$, and let ${\cal T}$ denote the phylogenetic tree with its branch lengths.
Let ${\cal M(\boldsymbol{\theta}, \boldsymbol{\phi})}$ denote the MMM that models the evolutionary process for all sites, where $\boldsymbol{\theta} = \{\boldsymbol \theta_1, \ldots, \boldsymbol \theta_{K}\}$ and $\boldsymbol\theta_k$ represents parameters for the $k$th evolutionary model, and $\boldsymbol{\phi}$ parameters of the switching process.
The observed data likelihood is:
\begin{equation}
L(\boldsymbol{\theta}, \boldsymbol{\phi}, {\cal T}, {\cal M} \mid \mathbf{Y}) = \prod_i \left( \sum_{(k, s)} \pi_{k}\pi_s L_i^R((k,s), \boldsymbol{\theta}, \boldsymbol{\phi}, {\cal T}, {\cal M} \mid \mathbf{Y}_i) \right),
\end{equation}
where the product is taken over every site $i$ in the alignment, with each site assumed to evolve independently.
The sum over the compound states $(k, s)$ replaces the sum over the nucleotide characters $s$ that is performed for standard nucleotide substitution models \citep{gascuel2007modelling}.
Here, $L_i^R((k,s), \boldsymbol{\theta}, \boldsymbol{\phi}, {\cal T}, {\cal M} \mid \mathbf{Y})$ is the likelihood of the data at site $i$ under category $k$ and given that state $s$ is observed at site $i$ of the root node $R$.
We can generalize this notation as $L_i^v((k,s), \boldsymbol{\theta}, \boldsymbol{\phi}, {\cal T}, {\cal M} \mid \mathbf{Y})$ for node $v$ to express the partial likelihood of observing the characters at site $i$ in the extant sequences descending from $v$.
This notation can be shortened to $L_i^v(k,s)$ because $\boldsymbol{\theta}$, $\boldsymbol{\phi}$, ${\cal T}$, ${\cal M}$ and $\mathbf{Y}$ are the same for all sites and nodes.
Let $l$ and $r$ be the left and right descendants of $v$  and $t_{v}$ the length of the branch connecting $v$ to its parent.
Each partial likelihood is then defined as follows (taking into account that the evolutionary categories are unobservable; \citet{gascuel2007modelling}):\vspace{0.35cm}

$L_i^v(k, s) =
\left \{
  \begin{tabular}{l}
  1 if $v$ is a leaf with nucleotide character $s$ \\
  0 if $v$ is a leaf with nucleotide character different from $s$ \\
  $\left( \sum_{(k', s')} P_{(k,s)(k',s')}(t_{l})L_i^l(k', s') \right) \left( \sum_{(k', s')} P_{(k,s)(k',s')}(t_{r})L_i^r(k', s') \right)$ otherwise.
  \end{tabular}
  \right.$\vspace{0.35cm}

\noindent The substitution probabilities $P_{(k,s)(k',s')}(t)$ are computed using Equation \ref{eq:matrixexp}, which requires a diagonalisation of $\mmRateMatrix$ that has computational complexity $\order{\numModels^3 \numStates^3}$ \citep{pan1999complexity}, although lower complexity may be achieved depending on the Kronecker structure for $\boldsymbol{\Lambda}$ (but see the Discussion section).
Computing these probabilities for all $\order{\numTaxa}$ branches in the phylogeny therefore has a complexity of $\order{\numTaxa \numModels^3 \numStates^3}$.
Evaluating the $L$ site likelihoods through the tree-pruning (or peeling) algorithm \citep{Felsenstein1981} amounts to a complexity of $\order{\numTaxa L \numModels^2 \numStates^2}$.
Taken together, with a relatively small cost $\order{L}$ for taking logarithm of site likelihoods and summing over sites results in a computational complexity of $\order{\numTaxa \numModels^3 \numStates^3 + \numTaxa L \numModels^2 \numStates^2}$ for the log-likelihood of the observed data.
We confront these computationally demanding tasks by heavily parallelizing their computation using powerful multi- and many-core hardware through the BEAGLE library \citep{beagle3}.

\subsection{Models and priors}

We use the HKY \citep{HKY} and GTR \citep{Lanave1984,Tavare1986} nucleotide substitution models in our examples.
We adopt a Yule pure birth process \citep{Yule1924} tree prior on the speciation process for the isochronous data sets, with a normally distributed prior on the log-transformed birth rate with mean 1.0 and standard deviation 1.5.
We assume a constant population size coalescent tree prior for the heterochronous influenza data set, with a normal prior on the log-transformed population size with mean 1.0 and standard deviation 1.5.
For the HKY model, we assume a normally distributed prior on the log-transformed transition-transversion ratio with mean 1.0 and standard deviation 1.25.
For the GTR model, we assume diffuse gamma distributed priors on its relative rate parameters.  
In particular, we employ a gamma distribution with shape 0.05 and scale 20.0 for $r_{AG}$ and gamma distribution with shape 0.05 and scale 10.0 for $r_{AC}$, $r_{AT}$, $r_{CG}$ and $r_{GT}$.
We assume an exponential prior with mean 0.5 on the rate heterogeneity parameter of the ASRV model \citep{yang1994maximum} as well as on the parameter describing the discretized gamma distribution to generate relative rate multipliers $\rho_i$.
Additionally, we assume a CTMC reference prior on the strict molecular clock rate in the heterochronous data set \citep{ferreira08} and an exponential prior with mean 1/3 on the standard deviation of the relaxed clock model with an underlying lognormal distribution for the isochronous data sets.
Finally, we assume an exponential prior with mean 1.0 on each of the switching rate parameters $\phi_{ij}$ and a Dirichlet prior with concentration parameters
equal to 1.0 on each set of nucleotide base frequencies.
For all analyses performed in the results section, the effective sample size (ESS) values for all estimated parameters were well above 200, as identified using Tracer \citep{rambaut2018posterior}.

\subsection{Bayesian model comparison}

Bayesian phylogenetic inference focuses on the joint posterior probability distribution of the tree and all evolutionary model parameters conditional on the observed sequence data.
Employing Markov chain Monte Carlo (MCMC) methods to approximate this posterior distribution handily avoids computing the marginal likelihood $p(\mathbf{Y} \mid {\cal M})$ that is notoriously difficult to estimate.
In order to compare the relative performance of our proposed MMMs to standard nucleotide substitution models for phylogenetic inference, however, estimating the marginal likelihood remains essential.
In all of our examples, we employ generalized stepping-stone sampling (GSS; \citet{Fan,baele2016genealogical}), an estimation procedure that accommodates phylogenetic uncertainty and provides accurate marginal likelihood estimates in a feasible amount of time.
GSS requires working distributions \citep{Fan,baele2016genealogical} for all parameters based on the posterior samples collected when exploring the preceding Markov chain.
BEAST \citep{beastX} constructs these distributions using kernel density estimation (KDE) based on Gaussian kernels and Silverman's rule to determine the bandwidth \citep{Silverman}.
Using MMMs will often require estimating a large collection of parameters, which may lead to poor ESS values for those parameters and also a more challenging KDE.
In order to avoid manually inspecting every MCMC analysis prior to starting the marginal likelihood estimation process, we suggest using the prior (distribution) as the working distribution for the MMM parameters as a general approach in conjunction with the specific tree working priors proposed in \citet{baele2016genealogical}.

\section{Results}

We here consider substitution models that are time-reversible and therefore substitution model $k$ will have instantaneous rates $Q^{(k)}_{ij}$ that can be expressed in terms of base frequencies $\pi_{kj}$ and symmetric rate parameters $R^{(k)}_{i \leftrightarrow j} = R^{(k)}_{j \leftrightarrow i}$ as follows:
\begin{equation}
Q^{(k)}_{ij} = \pi_{kj} R^{(k)}_{i \leftrightarrow j}.
\end{equation}
Thus a substitution model can be specified in terms of its base frequencies and symmetric rate parameters $\textbf R_k = \{R^{(k)}_{i \leftrightarrow j} | i \neq j, (i,j) \in \mathcal S \} $.

We adopt the following notation: MMM($M$)$_{ijk}$, where $M$ denotes the type of substitution model and $i$, $j$, and $k$ denote the numbers of distinct sets of symmetric rate parameters, sets of base frequencies, and relative rate multipliers, respectively.
For example, an MMM(HKY)$_{222}$ refers to an MMM featuring two different HKY substitution models, each with its own set of symmetric rate parameters and set of base frequencies, and two different relative rate multipliers.
An MMM(HKY)$_{122}$ refers to an MMM featuring two different HKY substitution models that share the same set of symmetric rate parameters but have different sets of base frequencies, along with two different relative rate multipliers.
When the relative rate multipliers are all fixed to 1 to superimpose an ASRV model \citep{yang1994maximum,Yang1996}, the $k$ subscript is omitted (for example, MMM(HKY)$_{22}$).

\subsection{Bacterial 16S ribosomal RNA}

Differences in base composition throughout the genome can bias phylogenetic inference when not properly taken into account.
Often, the proportion of A+T in a genome differs from that of G+C, and different organisms exhibit different patterns of base composition. 
At the level of the entire genome, GC content varies greatly within and among major groups of organisms, which can skew phylogenetic reconstruction if not properly unaccounted for \citep{Mooers2000}.
Two different evolutionary processes have been singled out as possible explanations for varying patterns of base composition: biases in the underlying process of mutation, as similar levels of GC content are often found in regions with different functional constraints, and natural selection, with increased global GC content in bacteria possibly being selected for by UV exposure \citep{singer1970}.

Environmental variation shaping nucleotide composition may cause unrelated taxa to share similar base composition and therefore be grouped together within a clade.
In order to accurately reconstruct evolutionary histories through phylogenetic inference, these potentially differing base compositions need to be accommodated in an explicit manner by the nucleotide substitution model.
To address this, \citet{Blanquart2006} developed a non-stationary and non-homogeneous model accounting for compositional biases, allowing the composition to change at random points in the tree, with the total number of change points across the tree being inferred from the data.
Through a Bayesian analysis of eubacterial 16S rRNA and BAS1 gene yeast data sets, the authors show that in most cases, the stationarity assumption was rejected in favor of their non-stationary model.

We evaluate our MMM framework on 16S ribosomal RNA of five bacterial sequences: \emph{Deinococcus radiodurans}, \emph{Thermus thermophilus}, \emph{Thermotoga maritima}, \emph{Aquifex pyrophilus} and \emph{Bacillus subtilis} (GenBank accession numbers: Y11332.1, AJ251939.1, NR\_029163.1, M83548.2 and CP009796.1).
We use standard nucleotide substitution models as well as MMMs to infer their evolutionary history while fixing the \emph{Aquifex pyrophilus} sequence as an outgroup.
The true tree topology of this eubacterial data set is believed to group \emph{D.~radiodurans} and \emph{T.~thermophilus} together to the exclusion of \emph{B.~subtilis}, \emph{T.~maritima}, and \emph{A.~pyrophilus}, given that \emph{D.~radiodurans} and \emph{T.~thermophilus} share the same peptidoglycan and menaquinone type \citep{murray1991}.
However, phylogenetic reconstruction under stationary models has a tendency to erroneously group \emph{D.~radiodurans} and \emph{B.~subtilis} together, because these mesophiles have similar, relatively low GC content.

\begin{figure}[!htbp]
\begin{center}
\includegraphics[scale=0.575]{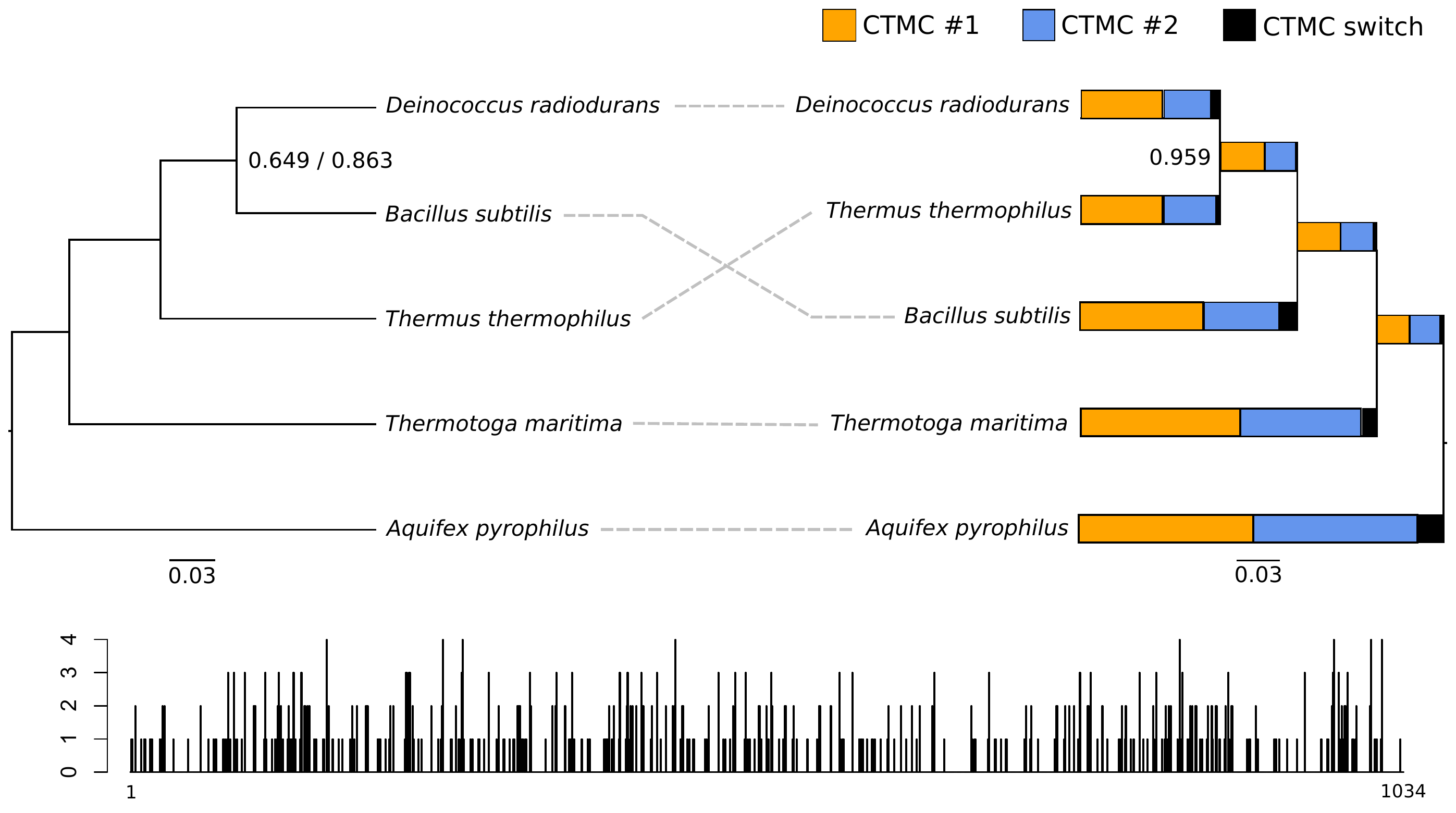} \\ \vspace{0.5cm}
\includegraphics[scale=0.45]{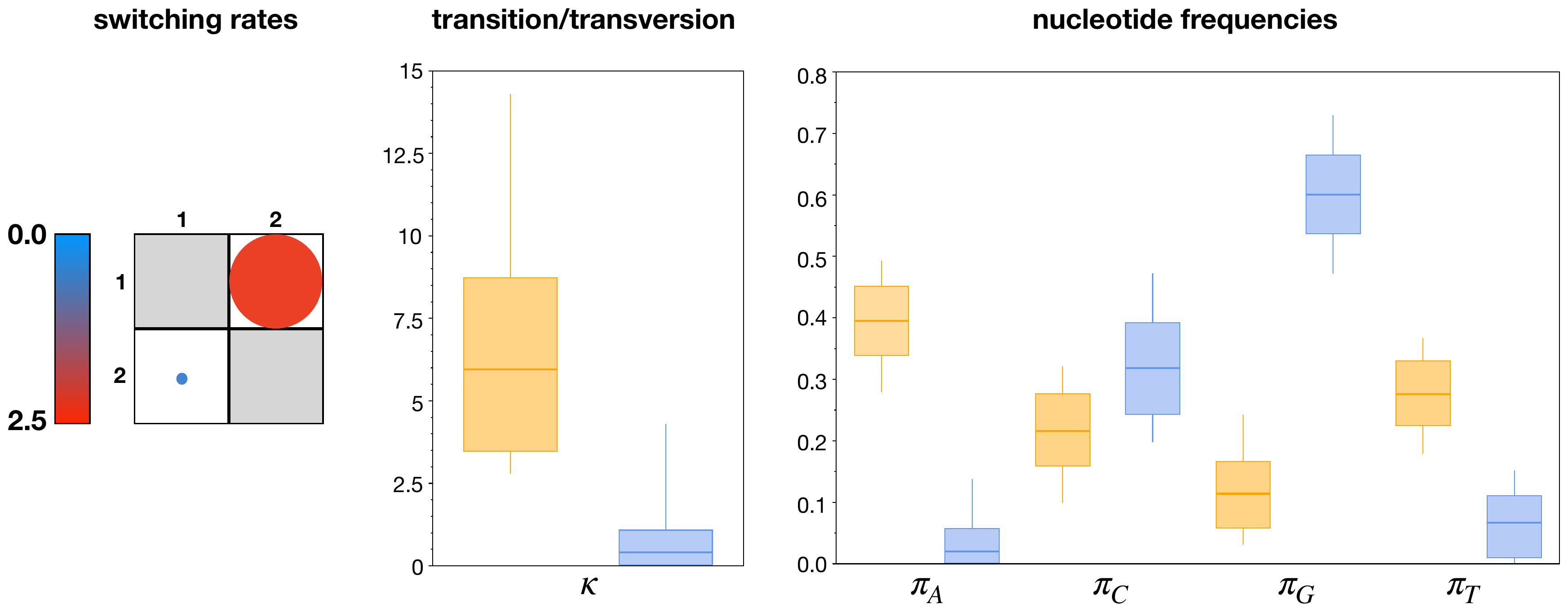}\vspace{-0.5cm}
\end{center}
\caption{Top: Maximum clade credibility (MCC) phylogeny relating five bacterial 16S sequences; unlabeled nodes have $>.9999$ posterior probability.
Standard nucleotide substitution models that assume among-site rate variation (ASRV) erroneously cluster the two mesophiles together with increasingly high posterior probability (0.649 for HKY and 0.863 for GTR in the topology on the left).
However, an MMM(HKY)$_{22}$ yields the correct clustering of the \emph{Deinococcus radiodurans} and the \emph{Thermus thermophilus} sequences with high posterior probability (topology on the right); each branch is annotated with the proportion of sites in each of the continuous-time Markov chain (CTMC) models, based on the maximum a posteriori (MAP) phylogeny.
Middle: number of CTMC model switches per alignment site based on the most probable hidden state realisations of the MMM on the MAP phylogeny; of the full alignment of 1304 sites, 761 sites are estimated not to switch between CTMC models. 
Bottom: mean posterior parameter estimates of the MMM show asymmetric switching between models with drastically different transition/transversion ratios and base frequencies.}
\label{fig:bacterial16S}
\end{figure}

Figure~\ref{fig:bacterial16S} shows the results of these phylogenetic reconstructions, with the HKY and GTR models -- both featuring an ASRV model and a relaxed molecular clock with an underlying lognormal distribution -- yielding similar (log) marginal likelihoods.
Note that, because we will include an ASRV model in all of these MMMs, we set all $\rho_{i}$ in equation~\ref{eq:singlematrix} to 1 to ensure identifiability.
Both the HKY and GTR models express strong support in favor of a clustering of \emph{D.~radiodurans} and \emph{B.~subtilis} (see Figure~\ref{fig:bacterial16S}), with the GTR model yielding a small increase in model fit to the data over the HKY model (log BF $<$ 1).
As such, both models yield an incorrect clustering, which appears to be primarily based on both sequences being mesophilic (low GC content), whereas the three other sequences are considered thermophilic (high GC content).
While an MMM of the type introduced by \citet{tuffley1998modeling} offers no improvement over these models when $\rateMatrix{1}$ is parameterised as an HKY model, a significant improvement in model fit can be obtained when $\rateMatrix{1}$ is parameterised as a GTR model (log BF = 19).
However, any MMM with two sets of base frequencies and with either a single set of symmetric rate parameters (an MMM($*$)$_{12}$) or with two different sets of symmetric rate parameters (an MMM($*$)$_{22}$) offers a further improvement in model fit compared to the standard nucleotide substitution models tested (25 $<$ log BF $<$ 35).
This can be attributed to the fact that MMMs are able to accommodate differing base compositions throughout the tree topology, and consequently yield an accurate phylogenetic reconstruction of the bacterial relationships, with the \emph{D.~radiodurans} and \emph{T.~thermophilus} clustering together (see Figure~\ref{fig:bacterial16S}) \citep{Embley1993,Mooers2000}.

\begin{figure}[!htbp]
\begin{center}
\includegraphics[scale=0.55]{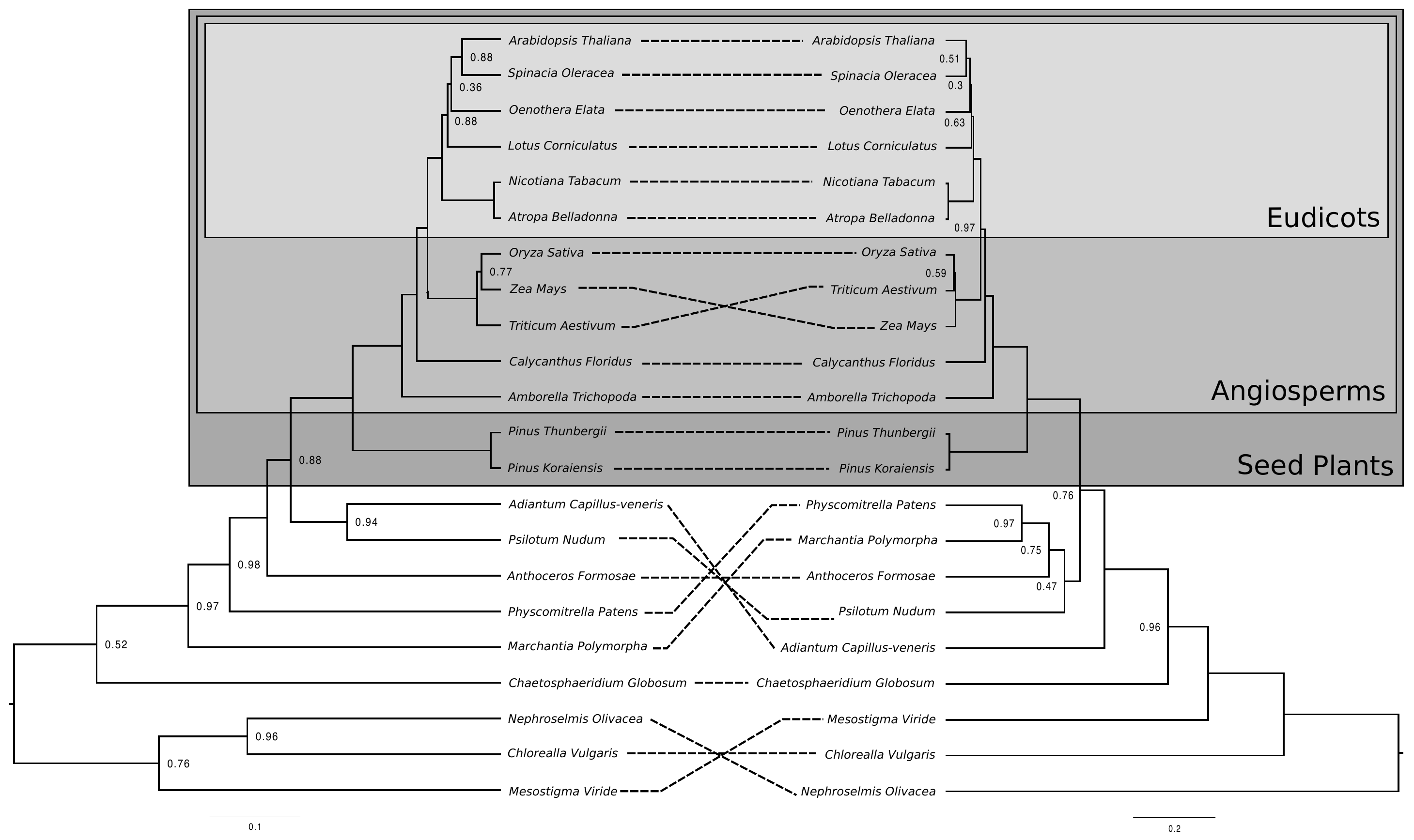}\vspace{-0.5cm}
\end{center}
\caption{Phylogenetic reconstruction of plant plastid sequences, based on the \psaB protein-coding gene; unlabeled nodes have $>.9999$ posterior probability.
Left: MCC tree based on a standard GTR model that features an ASRV model with 4 rate classes.
Right: MCC tree based on an MMM(GTR)$_{33}$ and featuring an ASRV model with 4 rate classes, which is strongly supported over the MCC tree generated under the GTR model (log Bayes factor of 356).
While only a single different clustering can be observed within the \emph{Angiosperms}, many differing clusters that have very high posterior probabilities are generated using the MMM(GTR)$_{33}$ outside of the seed plants.}
\label{fig:psaB}
\end{figure}

The base frequency estimates for the CTMC models within the MMM reflect the presence of mesophilic sequences (low GC content; orange in Figure~\ref{fig:bacterial16S}) and thermophilic sequences (high GC content; blue in Figure~\ref{fig:bacterial16S}) in our data, illustrating the capabilities of these MMMs.
Despite the fact that only eight branches connect the observed sequences, alignment sites switch up to four times between CTMC models across the phylogeny, indicating evolutionary patterns that cannot possibly be accommodated using standard nucleotide substitution models.
Over forty percent of the alignment sites undergoing at least one switch between CTMC models in a highly asymmetric manner (see Figure~\ref{fig:bacterial16S}), with the models also exhibiting drastically differing transition/transversion ratios.

In conclusion, we show that appropriately modeling compositional heterogeneity for these eubacterial sequences enables inference of the correct phylogeny as well as base frequency compositions that reflect the presence of both mesophilic and thermophilic sequences in the data set.
We note that similar observations regarding phylogenetic inference under compositional heterogeneity have been made in previous studies.
For example, \citet{Foster2004} used a heterogeneous model that accommodates compositional differences over the tree, without requiring that each branch is equipped with its own independent base composition, allowing modeling of compositional heterogeneity with few additional parameters.
This reflects the assumption that compositional differences do not vary continuously over the tree.
The compositional heterogeneity model of \citet{Foster2004} yields the correct topology, as does the non-stationary and non-homogeneous model of \citet{Blanquart2006}, while standard nucleotide substitution models fail to do so.

\subsection{Plant plastid genes}
\label{sec:plastid}

We consider nucleotide sequence data from the protein-coding genes of 23 completely sequenced plant plastid genomes, previously analysed by \citet{ane2005} to measure the independence of the substitution process between two groups of taxa as a means of detecting covarion evolution.
Assuming a fixed underlying reference tree that represents the likely relationships of plant taxa for which complete chloroplast sequences were available at the time, the covarion test of \citet{ane2005} detected significant covarion evolution ($P < 0.0005$) in 14 of 57 genes analysed across all positions.
We here analyse the \emph{psaB} and \emph{ndhD} genes with standard nucleotide substitution models and MMMs and compare the inferred phylogenies and model fit.

A comparison of standard nucleotide substitution models reveals that the combination of a GTR model and an ASRV model, along with a relaxed clock assuming an underlying lognormal distribution, yields the highest (log) marginal likelihood for both data sets.
We conduct analyses with MMMs that feature an HKY or GTR substitution model with a single set of symmetric rate parameters along with two or three different sets of base frequencies (i.e. MMM($*$)$_{12}$ and MMM($*$)$_{13}$ models), as well as generalizations of these MMMs that feature as many different sets of symmetric rate parameters as sets of base frequencies (i.e. MMM($*$)$_{22}$ and MMM($*$)$_{33}$ models).
For all of these models, we set all $\rho_{i}$ in equation~\ref{eq:singlematrix} to 1 to ensure identifiability when using an ASRV model in combination with MMMs.
We also analyse the data with a nucleotide covarion model \citep{tuffley1998modeling}, which we can easily compose within our MMM framework through XML specification.

For the \emph{psaB} data set, the covarion-style model is strongly preferred over a standard GTR+ASRV substitution model by a log Bayes factor of 208.
The MMM(GTR)$_{12}$ and MMM(GTR)$_{22}$ yield log Bayes factors of 63 and 316, respectively, over the standard GTR+ASRV model.
MMM(GTR)$_{13}$ and MMM(GTR)$_{33}$ parameterisations yield further increases in model fit of 327 and 356, respectively, over the GTR+ASRV model.
Because additional categories within the MMM offer diminishing returns in terms of model fit at the expense of additional computation time, we did not explore MMMs with even higher dimensions.
Figure~\ref{fig:psaB} shows the maximum clade credibility (MCC) trees obtained under the standard GTR+ASRV model and the MMM(GTR)$_{33}$ that generated the highest (log) marginal likelihood.
While the clustering within the seed plants is identical under both models, substantial differences in posterior support can be observed for specific clades.
In the remaining part of the tree, these models result in completely different clustering patterns with strong support for many clades under the MMM(GTR)$_{33}$ model.

\begin{figure}[!htbp]
\begin{center}
\includegraphics[scale=0.525]{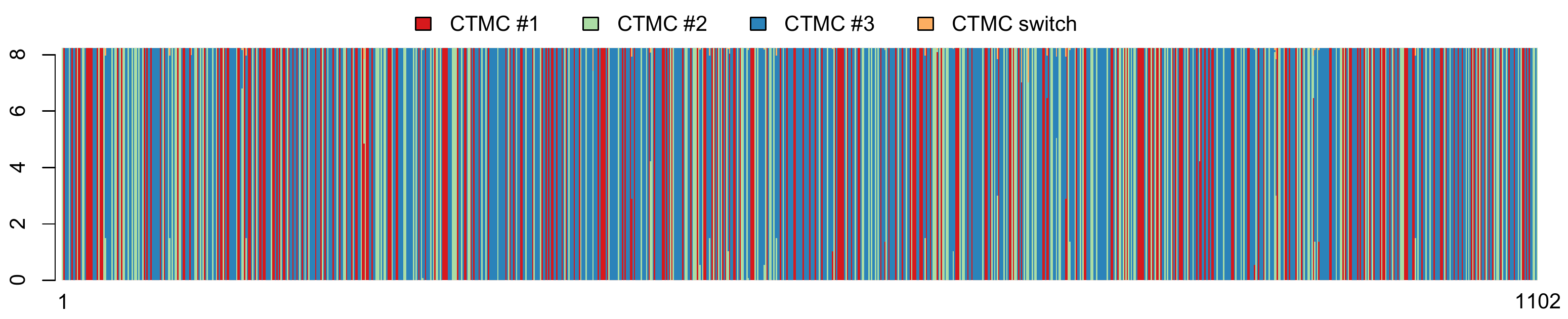}
\includegraphics[scale=0.525]{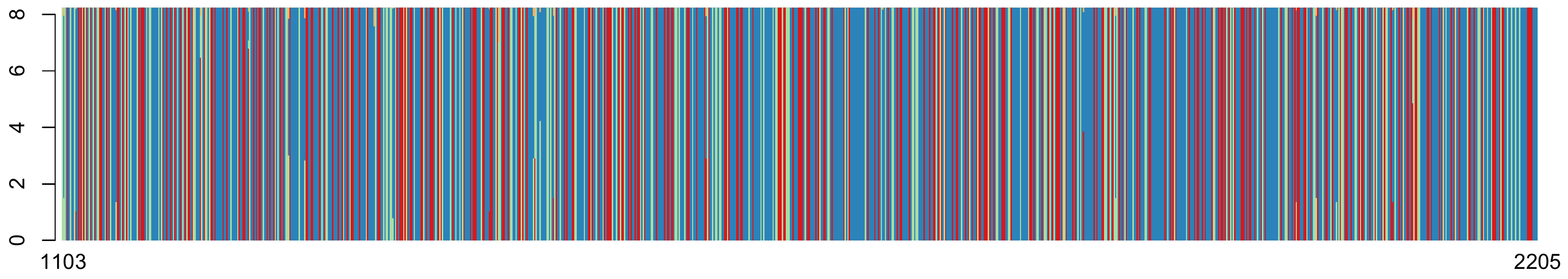}\vspace{-0.25cm}
\begin{minipage}[t]{0.4\textwidth}
\vspace{0pt}
\includegraphics[scale=0.58]{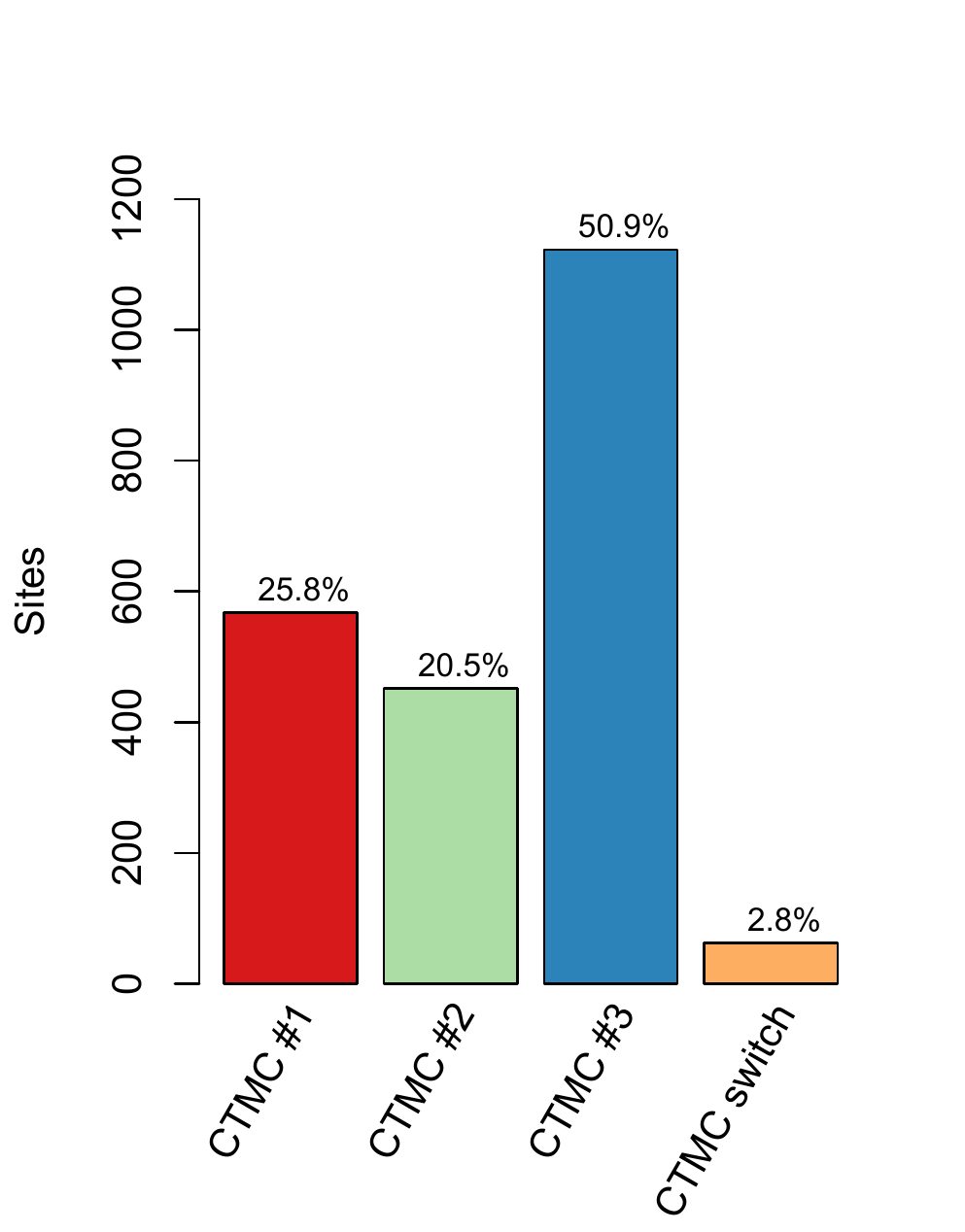}
\end{minipage}
\begin{minipage}[t]{0.6\textwidth}
\vspace{0pt}
\includegraphics[scale=0.60]{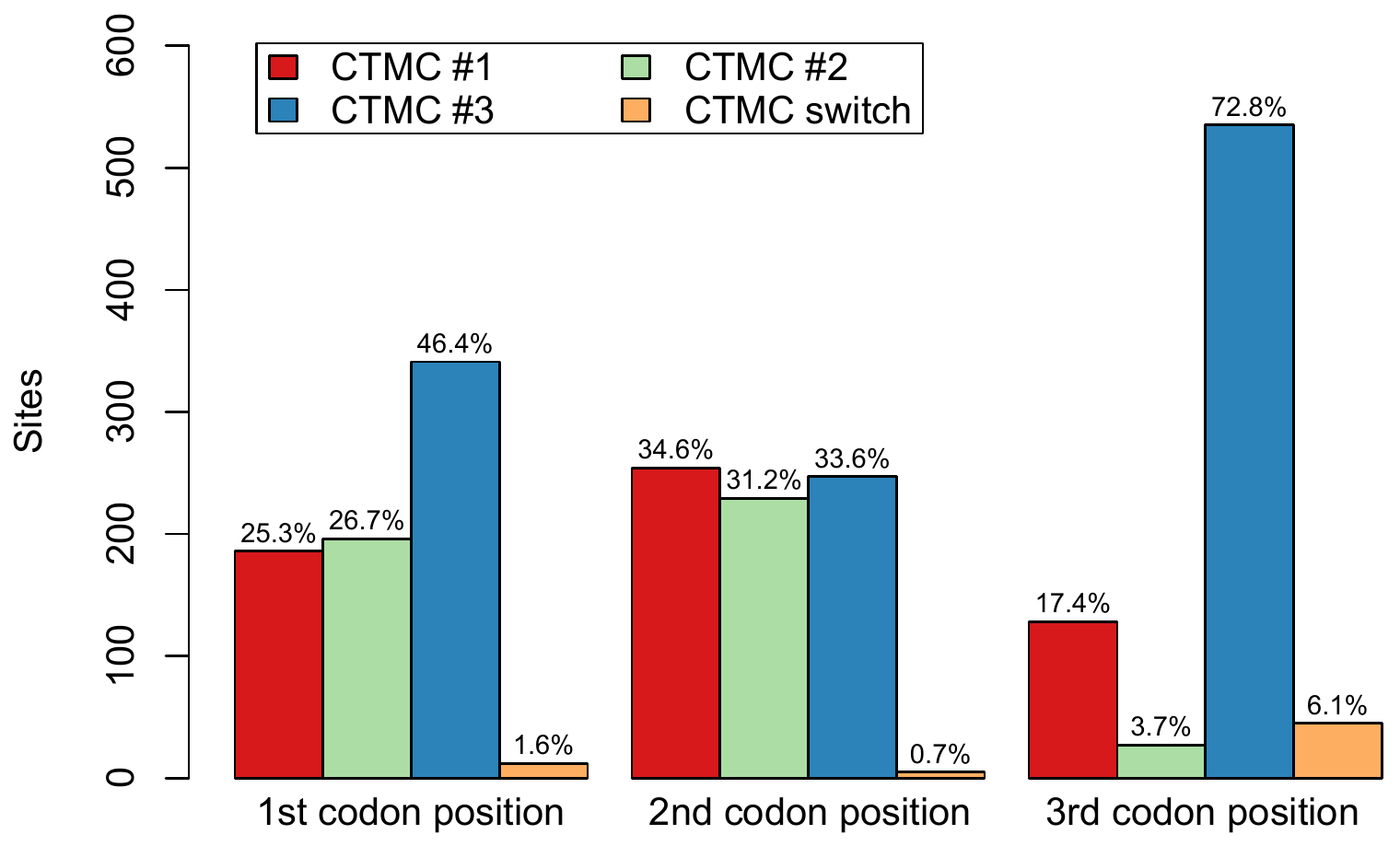}
\end{minipage}
\includegraphics[scale=0.45]{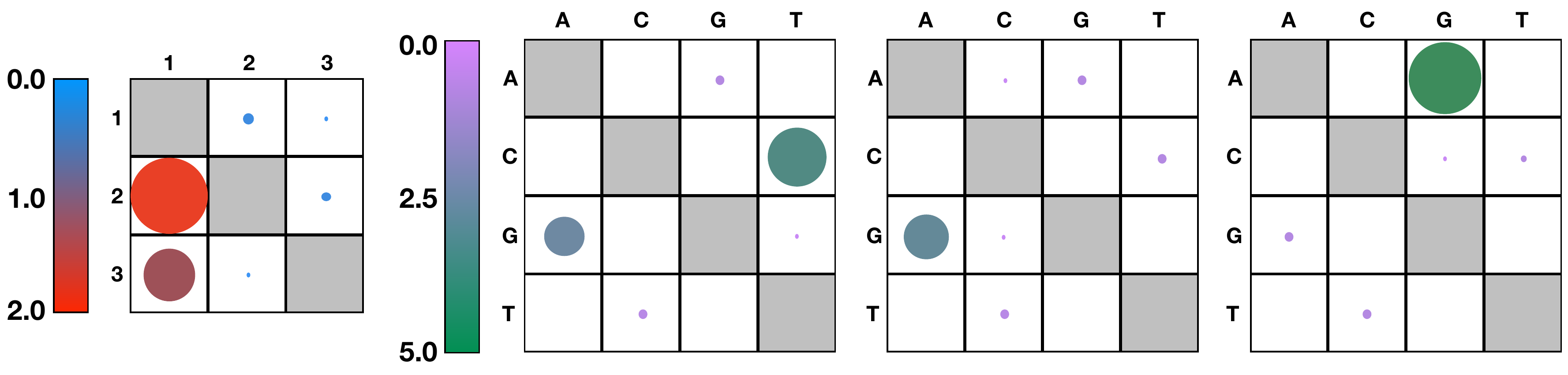}
\end{center}\vspace{-0.60cm}
\caption{Markov-modulated model behaviour on the \psaB protein-coding gene phylogeny.
Top: amount of time (branch lengths in genetic distance) spent in each CTMC model for each alignment site based on the most probable hidden state realisations of the MMM on the maximum a posteriori phylogeny.
Middle left: summary of the number of sites that evolve according to each CTMC, illustrating complex substitution patterns that go beyond codon position partitioning, as well as 2.8\% of sites switching between CTMC realisations.
Middle right: distribution of sites in each codon position across the different CTMC model realisations, showing that first and second codon positions switch far less frequently between CTMC models than the third codon position, in which the substitutions occur according to a clearly predominant CTMC.
Bottom: switching behaviour of the MMM between the three CTMC models, with the main instantaneous substitution rates shown for those models.}
\label{fig:psaB-mmm}
\end{figure}

In Figure~\ref{fig:psaB-mmm}, we illustrate the complex substitution patterns across all sites on the maximum a posteriori (MAP) \psaB phylogeny, using the most probable hidden state realisations of the MMM(GTR)$_{33}$.
We use a simple counting procedure to reconstruct which sites evolve according to which CTMC within the MMM(GTR)$_{33}$, and we observe a relatively small amount of CTMC switching throughout the phylogeny.
The reconstructed patterns go beyond mere codon position partitioning, as we observe different substitution dynamics per codon position. 
In particular, the third codon position is the only position that evolves according to a particular CTMC a majority of the time, and it also exhibits the greatest degree of switching between CTMC realisations. 
We also depict the mean posterior instantaneous substitution rates of the various MMM components in Figure~\ref{fig:psaB-mmm}, showing a clearly asymmetric CTMC switching process and three distinct GTR model realisations within the MMM.
This complex interplay of model components is consistent with the strong Bayes factor support of the MMM(GTR)$_{33}$ over all other models tested.

\begin{figure}[!htbp]
\begin{center}
\includegraphics[scale=0.55]{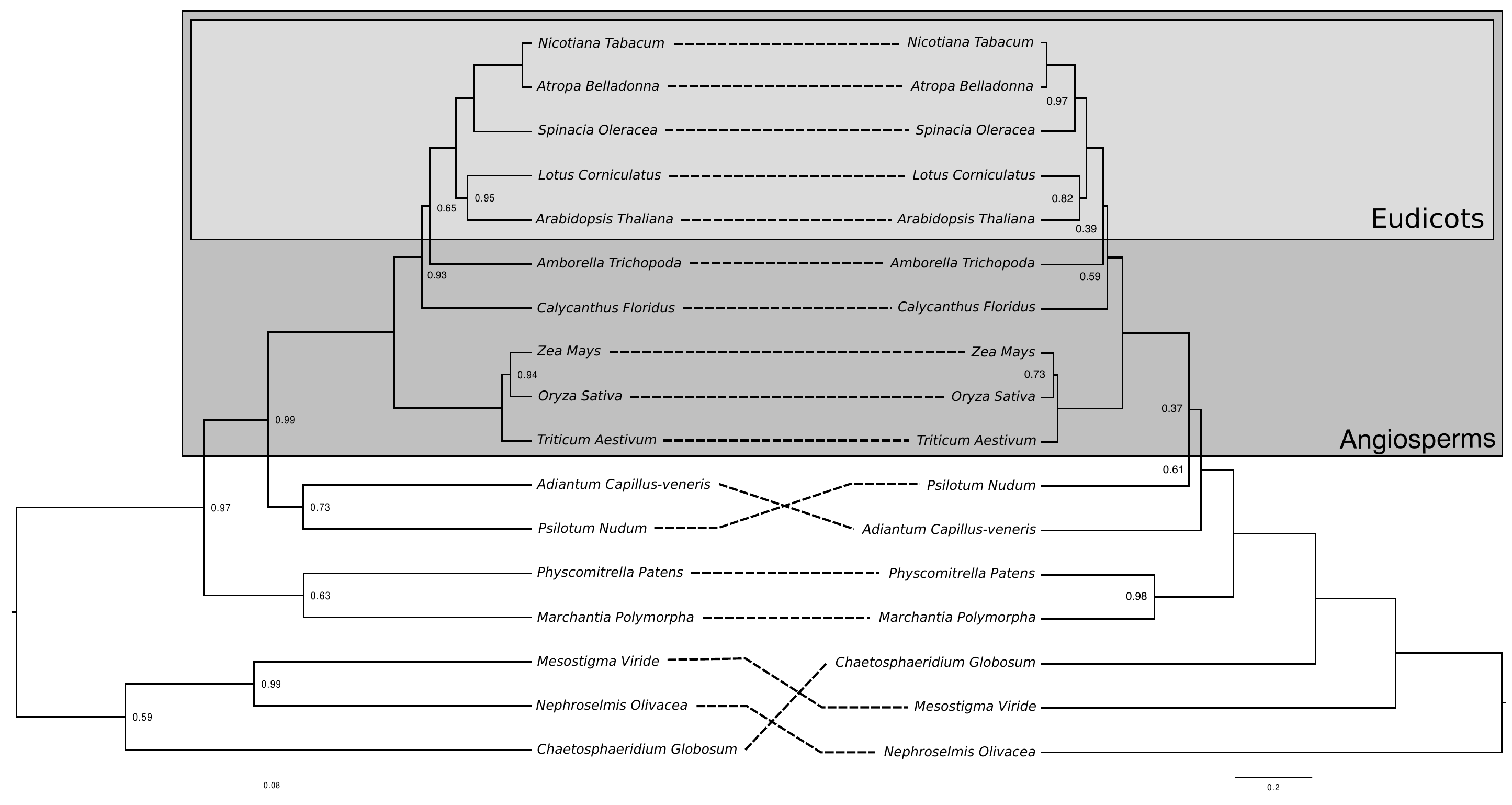}\vspace{-0.5cm}
\end{center}
\caption{Phylogenetic reconstruction of plant plastid sequences, based on the \ndhD protein-coding gene; unlabeled nodes have $>.9999$ posterior probability.
Left: MCC tree based on a standard GTR model that features an ASRV model with 4 rate classes.
Right: MCC tree based on an MMM(GTR)$_{33}$ and features an ASRV model with 4 rate classes, which is strongly supported over the MCC tree generated under the GTR model (log Bayes factor of 184).
The differences in clustering are all situated outside of the \emph{Angiosperms} clade.
}
\label{fig:ndhD}
\end{figure}

For the \emph{ndhD} data set, the covarion-style model is strongly preferred over a standard GTR+ASRV substitution model, yielding a log Bayes factor of 84 in favor of the former.
However, the MMM(GTR)$_{12}$ and MMM(GTR)$_{22}$ yield a log Bayes factor of respectively 138 and 155 over the standard GTR+G model.
Again, further increases in model fit can again be obtained using the MMM(GTR)$_{13}$ and MMM(GTR)$_{33}$ parameterisations, of respectively 171 and 184 over the GTR+ASRV model.
We deemed the performance increase of the MMM(GTR)$_{33}$ over the MMM(GTR)$_{22}$ insufficient to warrant testing even higher-dimensional MMMs.
Figure~\ref{fig:ndhD} shows the inferred phylogenies for the standard GTR+ASRV model and the MMM(GTR)$_{33}$ that generated the highest (log) marginal likelihood.
The clustering within the angiosperms is identical under both models, but substantial differences can again be observed in the posterior support of particular clades.
Basal to the angiosperm clade, the clustering of \emph{Physcomitrella patens} with \emph{Marchantia polymorpha} is retrieved under both models but other differences can clearly be observed, often with strong posterior support under the MMM(GTR)$_{33}$.

\begin{figure}[!htbp]
\begin{center}
\includegraphics[scale=0.525]{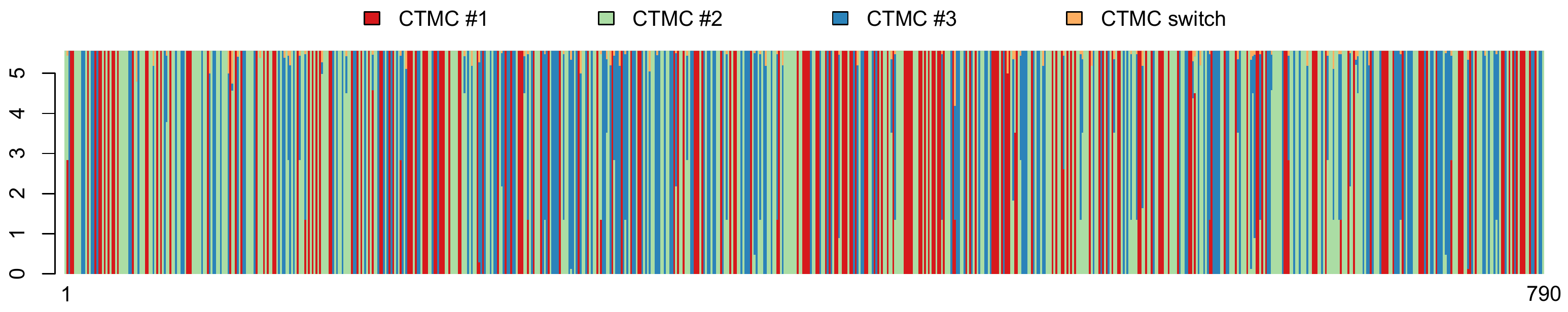}
\includegraphics[scale=0.525]{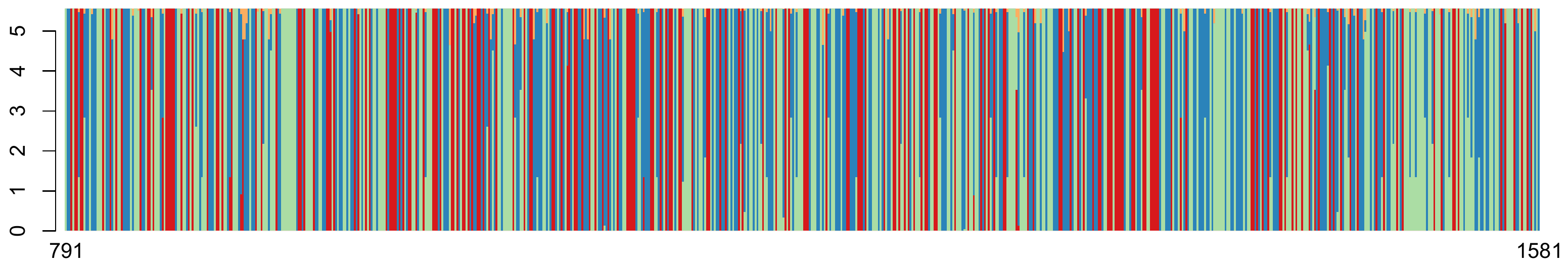}\vspace{-0.25cm}
\begin{minipage}[t]{0.4\textwidth}
\vspace{0pt}
\includegraphics[scale=0.58]{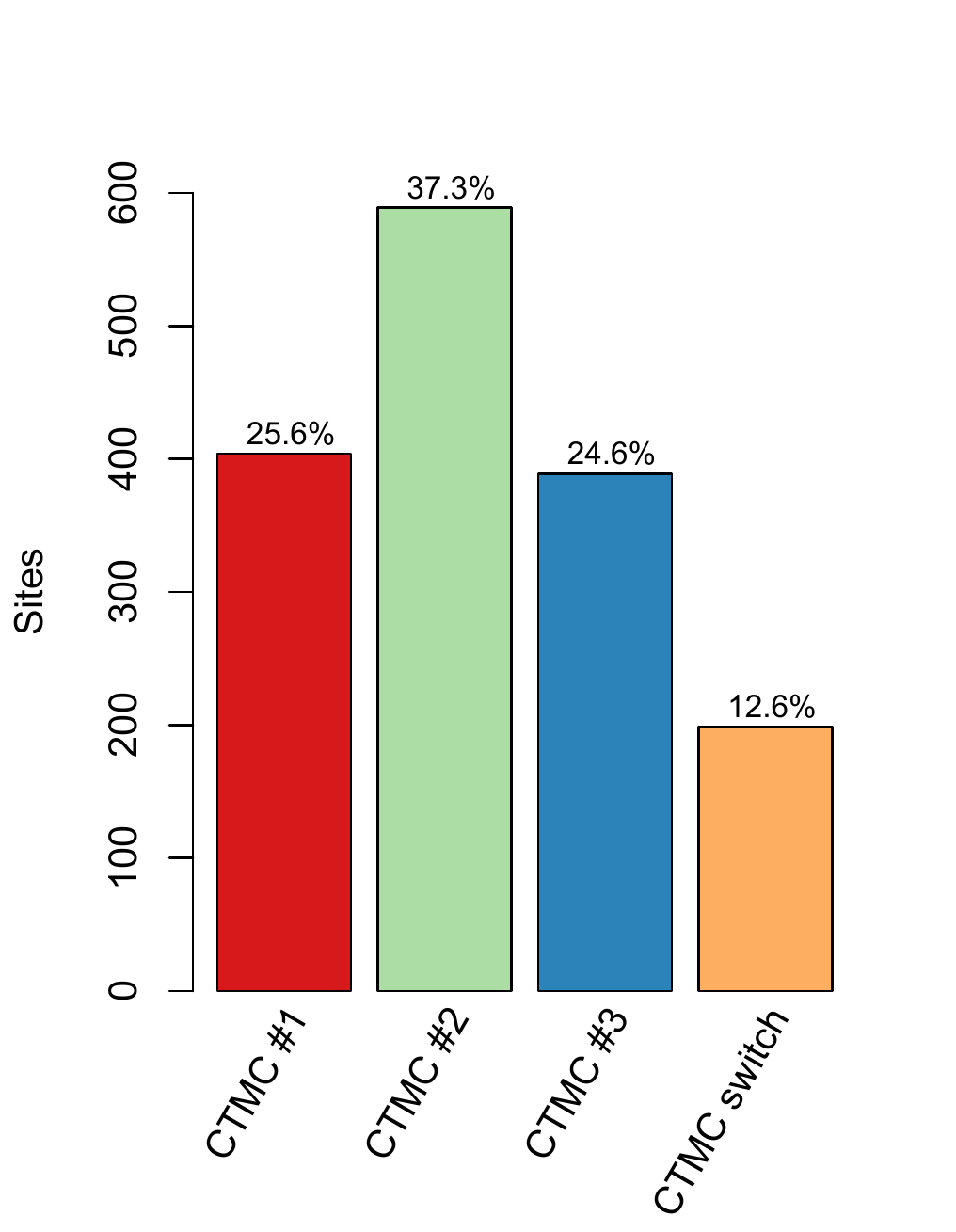}
\end{minipage}
\begin{minipage}[t]{0.6\textwidth}
\vspace{0pt}
\includegraphics[scale=0.60]{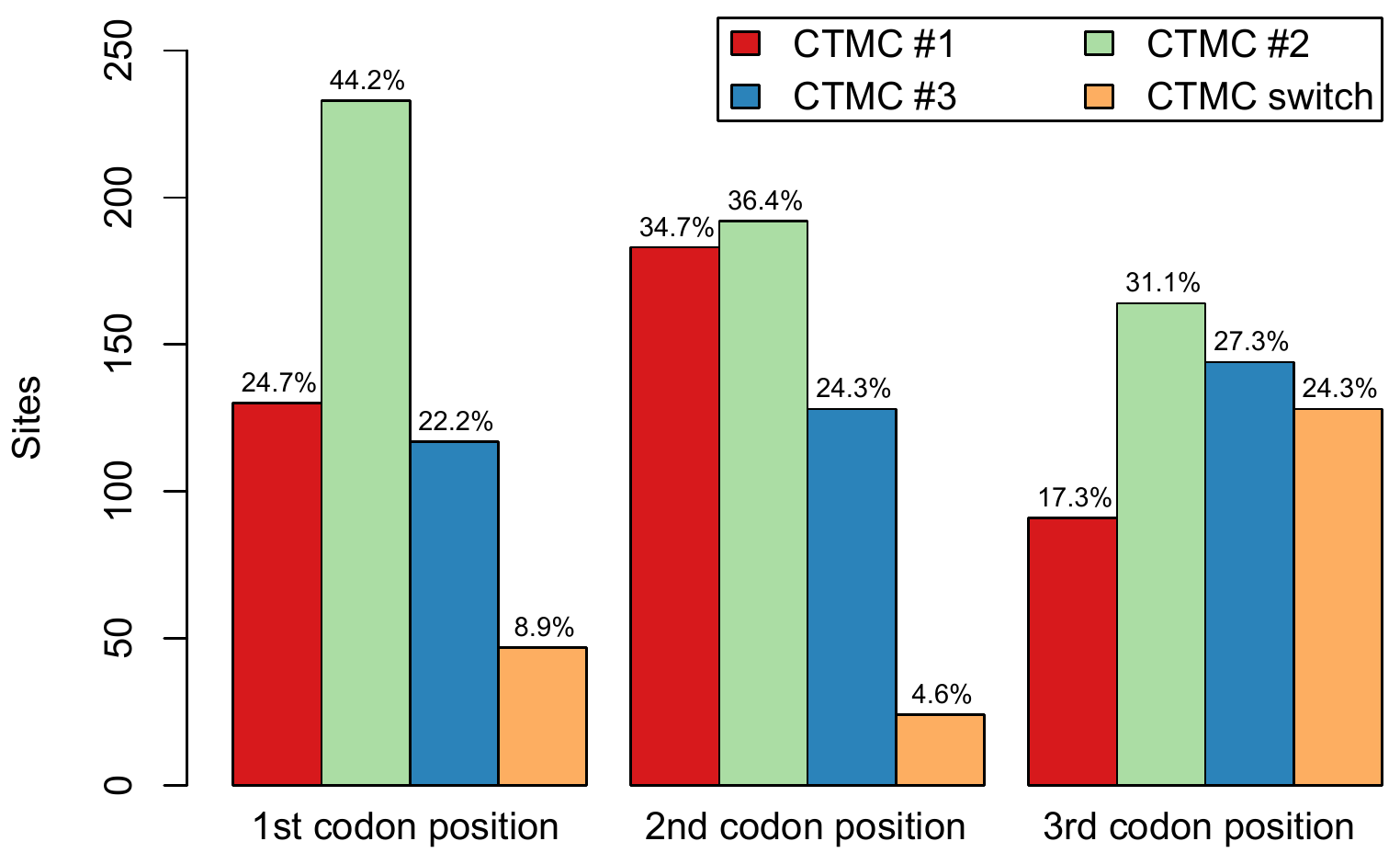}
\end{minipage}
\includegraphics[scale=0.45]{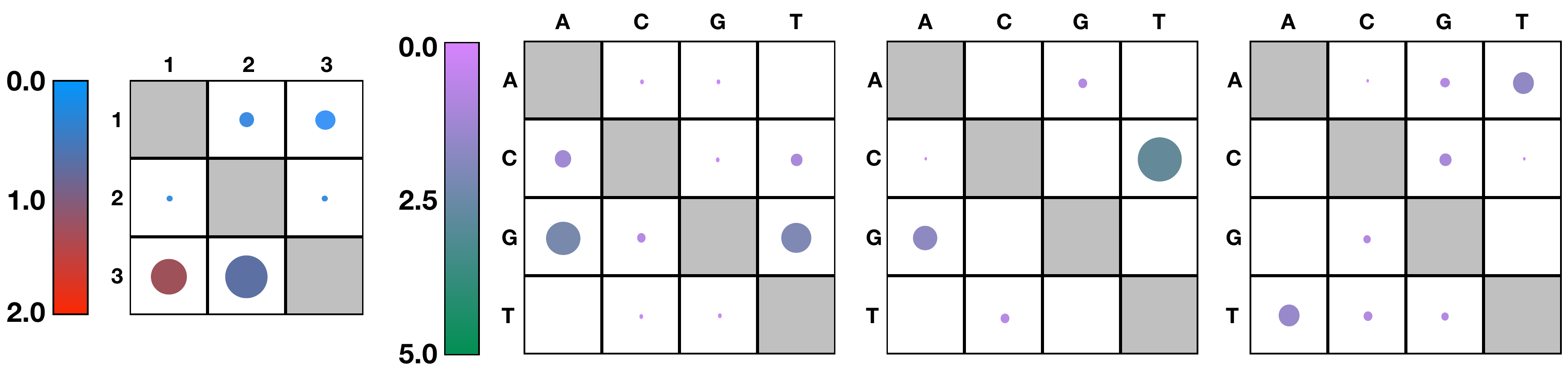}
\end{center}\vspace{-0.60cm}
\caption{Markov-modulated model behaviour on the \ndhD protein-coding gene phylogeny.
Top: amount of time (branch lengths in genetic distance) spent in each CTMC model for each alignment site based on the most probable hidden state realisations of the MMM on the MAP phylogeny.
Middle left: summary of the number of sites that evolve according to each CTMC, illustrating a large proportion (12.6\%) of sites switching between CTMC realisations.
Middle right: distribution of sites in each codon position across the different CTMC model realisations, again showing that first and second codon positions switch far less frequent between CTMC models than the third codon position, in which each CTMC describes the substitution patterns of a fair proportion of sites across the phylogeny.
Bottom: switching behaviour of the MMM between the three CTMC models, with the main instantaneous substitution rates shown for those models.}
\label{fig:ndhD-mmm}
\end{figure}

In Figure~\ref{fig:ndhD-mmm}, we illustrate the complex substitution patterns across all sites on the maximum a posteriori (MAP) \ndhD phylogeny, using the most probable hidden state realisations of the MMM(GTR)$_{33}$.
We again use a simple counting procedure to reconstruct per-site evolutionary patterns within the MMM(GTR)$_{33}$, and we see that a more substantial amount of CTMC switching occurs throughout the phylogeny for all positions than for the \psaB data set.   
The third codon position exhibits the greatest degree of switching between CTMC realisations, as with the \psaB data set. 
On the other hand, the first codon position in this \ndhD data set evolves according to a predominant CTMC, rather than the third codon position, as in the \psaB data set.
Given the substantial amount of CTMC switching across the \ndhD phylogeny, models that assume a fixed CTMC per position cannot accurately describe the complex substitution dynamics we observe here, explaining the large Bayes factors in favor of MMMs over traditional substitution models.
The mean posterior instantaneous substitution rates of the various MMM components show a less asymmetric CTMC switching process than for the \psaB data set, with again three clearly distinct GTR model realisations within the MMM.

In conclusion, we observe that MMMs infer substantially different phylogenies than standard nucleotide substitution models, and the differences yielded by MMMs are supported by significant increases in model fit.
In order to show that these large differences are not artifacts of using such high-dimensional models, we perform a small simulation study to assess the ability of MMMs to retrieve the generative models of simulated sequence alignments and to quantify their increase in model fit when the MMM was the generative model (Appendix A).
Our simulation study shows similar differences in model fit to the ones obtained in this section for the \emph{psaB} and \emph{ndhD} genes, as well as the ability of state-of-the-art Bayesian model selection to select the generative substitution model even when compared with similar model parameterisations.
Importantly, when simulating data under a standard GTR model, MMMs exhibit a worse model fit than the GTR model that generated the data.
These analyses of simulated data show that MMMs can easily be used in combination with recent developments in Bayesian model selection, and provide additional support for our conclusions that these models can yield substantial increases in model fit over standard nucleotide substitution models.

We note that each additional CTMC within an MMM leads to (much) increased computational demands, and that a search for the optimal MMM may hence prove time-consuming for complex large data sets.
In order to make such computations manageable, BEAST can however exploit the BEAGLE library \citep{beagle3} to offload the large matrix exponentiations and multiplications onto powerful multi-core hardware solutions.
In particular, the use of graphics cards for scientific computing yields significant performance gains over standard multi-core processors, rendering phylogenetic inference under these MMMs feasible despite their complexity (Appendix B).

\subsection{Assessing heterotachy using Markov-modulated models}

Phylogenetic inference using molecular clock models has proven to be crucial for accurate reconstruction of the origin and spread influenza A virus (IAV) within and between hosts \citep{rambaut2008}.
Recently, \citet{Worobey2014} showed that estimates under the popular uncorrelated relaxed clock model can be flawed when discrete rate variation exists among clades.
In a similar fashion, traditional substitution models may not be able to fully capture complex substitution dynamics over time that lead to non-stationary base distributions.
We here analyse a sequence alignment comprising the external antigenic haemagglutinin (HA) gene \citep{Worobey2014}, 
encompassing 454 taxa and spanning 1815 base pairs, providing for a computationally demanding analysis using MMMs, to test if the evolutionary rate of sites varies over time, a process known as heterotachy \citep{Lopez2002}.

We consider an MMM(HKY)$_{114}$ for this analysis, as we specifically focus on the patterns of evolutionary rate variation across the phylogeny, i.e. we impose
\begin{equation}
	\rateMatrix{\modelIdx} = \rateMatrix{}, \modelIdx = 1, \ldots, 4.
\end{equation}
We draw the relative rate multipliers ($\rho_k, \text{with } k = 1, \ldots, 4$) from a discretized gamma distribution $\Gamma(\alpha)$, with its shape $\alpha$ estimated from the data and its scale $\beta = \alpha$.
Note that this discretization is identical to how among-site rate variation is typically modelled \citep{yang1994maximum,Yang1996}.
We compare the performance of this MMM(HKY)$_{114}$ to that of a standard HKY substitution model (with and without an ASRV model).
We also assess the performance of partitioning according to codon position for each of the models tested.

\begin{figure}[!htbp]
\begin{center}
\includegraphics[scale=0.475]{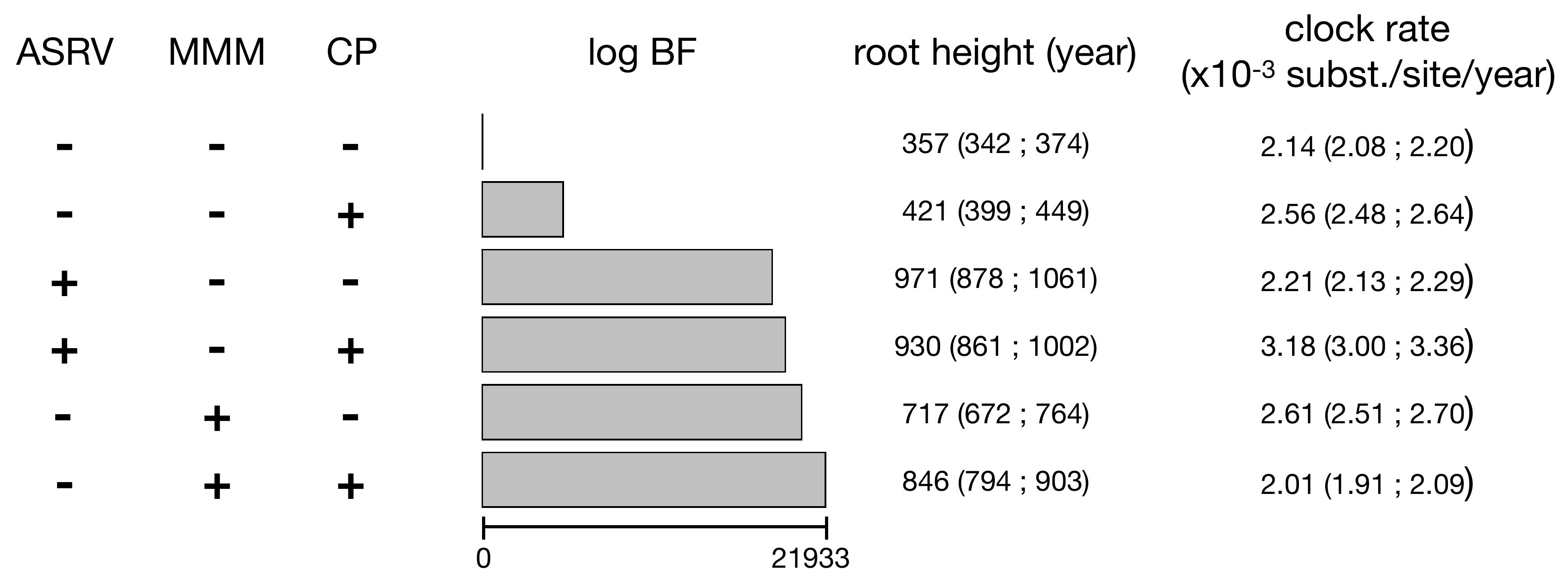}\\ \vspace{0.5cm}
\includegraphics[scale=0.45]{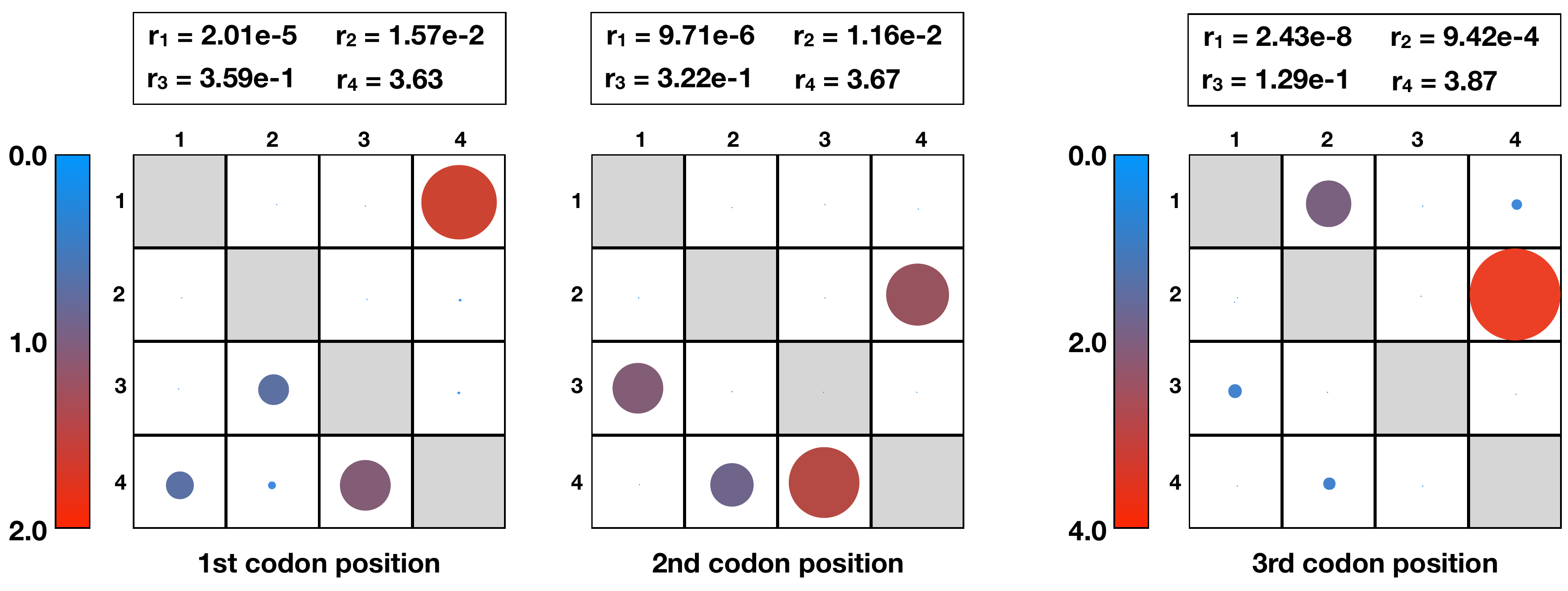} \vspace{-0.5cm}
\end{center}
\caption{Bayesian model comparison on the antigenic haemagglutinin data set.
Top: relative performance of MMMs compared to ASRV models, with and without codon position partitioning (CP).
MMMs substantially outperform ASRV models, with a log Bayes (BF) factor of 1841 when CP is ignored and a log BF of 2610 when CP is taken into account.
Accounting for ASRV results in much older divergence times compared to not making such an assumption, regardless of codon position partitioning.
MMMs yield more recent divergence time estimates than ASRV models, although the effect is somewhat mitigated by assuming a different MMM per codon position.
Bottom: posterior mean switching rate parameter estimates for the MMM+CP model, which attained the highest (log) marginal likelihood, illustrating different rate distributions (using 4 rate categories, with the rates shown in the rectangles) and rate switching dynamics for each codon position.}
\label{fig:HA}
\end{figure}

Figure~\ref{fig:HA} shows the relative performance of the different models tested on this data set along with their clock rate and root height estimates (95\% credibility intervals are shown in brackets).
As expected, accommodating ASRV substantially improves model fit for a standard nucleotide substitution model (with and without codon position partitioning), and also has a significant impact on root height estimation.
Partitioning according to codon position consistently improves model fit as well, as this has clear biological motivation \citep{Shapiro2006}.
Modeling heterotachy using an MMM(HKY)$_{114}$ yields further substantial improvements in model fit over ASRV models, both with and without codon position partitioning (log Bayes factors of 1841 and 2610, respectively), and results in a root height that is much older than the one estimated using a standard nucleotide substitution model without an ASRV model.
We did not, however, uncover systematic differences in the clock rate estimation with MMMs compared to standard substitution models (with and without an ASRV model).

We observe heterogeneous rate switching behaviour across codon positions for the top-performing MMM+CP model (Figure~\ref{fig:HA}).
This is consistent with the Bayes factor support of MMMs over ASRV models and illustrates that the assumption of constant rates for sites across the underlying phylogeny is an oversimplification.
We specifically observe strong asymmetric rate switching at the third codon position, indicative of reduced functional constraints at these positions.
These results shows the flexibility of our MMM implementation as well as the potential for MMMs to capture complex processes, such as heterotachy, that cannot be fully captured by popular models that are in frequent use today.
Finally, these analyses also show that despite the large number of sequences in this HA data set, Bayesian inference with MMMs -- with accompanying (log) marginal likelihood estimation -- is feasible on large data sets when employing a high-performance computing library on GPU \citep{beagle3}.

\section{Discussion}

Markov-modulated rate processes were originally developed for information-handling purposes in communication and computer systems, where the arrival rate of information fluctuates over time \citep{stern1991analysis}.
Consisting of a superposition of a number of independent finite state reversible Markov processes, the Markov-modulated Poisson process has seen extensive use in many applications (see e.g. \citet{fischer1993markov}), before finding its way into phylogenetics (see the introduction for an extensive overview).
Given the high dimensionality of MMMs, initial applications in phylogenetics focused on efficient approaches for MMM matrix diagonalization and exponentiation \citep{galtier2001maximum,galtier2004markov} and targeted MMMs with only a few free parameters \citep{guindon2004modeling}.

Specifically, when an MMM comprises evolutionary models that share the same instantaneous rate matrix (i.e. $\rateMatrix{\modelIdx} = \textbf Q$ for all $k$), the computational complexity of instantaneous rate matrix diagonalization (required for calculation of the data likelihood) is reduced via an efficient Kronecker formulation from $\order{\numStates^3 \numModels^3}$ to $\order{\numStates^3} + \order{\numStates \numModels^3}$, with $\numStates$ ranging from 4 for nucleotides, over 20 for amino acids, to 61 for codons \citep{galtier2004markov}.
However, this gain in computational efficiency is not applicable in the case of more than one unique $\rateMatrix{\modelIdx}$, which was the case in all of our examples, with the exception of the heterotachy test for the antigenic haemagglutinin example.
For such cases, more general matrix exponentiation via eigenvalue decomposition is required, for which the computational demands increase tremendously with the MMM's dimension.
However, substantial performance improvements in computational hardware now enable the use of MMMs featuring dozens of estimable parameters. In particular, high-performance computational libraries can alleviate the computational burden through offloading the most demanding routines to powerful graphics cards \citep{beagle3}.
Efficient scaling of higher dimensional MMMs may also be aided by innovations in MCMC sampling.
For example, adaptive MCMC transition kernels may allow for more efficient posterior exploration for parameter-rich MMMs, as such kernels have been shown to substantially increase ESS values for standard nucleotide substitution model parameters \citep{Baele2017}.

We have shown that MMMs can offer great improvements in model fit for a wide range of scenarios, from bacterial and viral to plant plastid data sets.
Notably, we demonstrated that MMMs can outperform the ASRV model, which has become ubiquitous in phylogenetic reconstruction \citep{yang1994maximum,Yang1996}.
An MMM is in fact a generalization of the ASRV model, made possible via constructing $\mmRateMatrix$ directly from scaled versions of $\rateMatrix{\modelIdx}$ (as we have shown in this manuscript) while also allowing for branch rate variation.
We have employed the same discretization of an underlying gamma distribution for determining the relative rate multipliers of the $\rateMatrix{\modelIdx}$, although other approaches may also prove useful towards this end.

A major strength of our framework is its generality. Researchers can compose a wide variety of MMMs through a flexible XML specification for analyses in BEAST.
An MMM can comprise an arbitrary number of evolutionary models that can differ in relative substitution rates, relative character exchange rates, and stationary distributions.
However, determining the ideal composition of an MMM can be challenging. It may not be obvious how many and which evolutionary models to use as its building blocks.
To mitigate the need for such difficult \textit{a priori} decisions and to avoid the pitfalls of adopting unsuitable models, we plan on employing Bayesian nonparametrics to infer the number of evolutionary models \citep{Beal2002} and Bayesian model averaging techniques to account for evolutionary model uncertainty \citep{Wu2013}.

Most of the development of MMMs in evolutionary inference has focused on their utility in phylogenetic reconstruction from molecular sequence data.
However, there is also a need for MMMs to analyse phenotypic traits.
For example, \citet{beaulieu2013identifying} investigated the evolution of growth habit in the angiosperm clade \emph{Campanulidae} by employing a covarion-like model that modulates between ``fast'' and ``slow'' hidden regimes.
We anticipate that the more general MMM formulation we consider here will also prove useful in the context of phenotypic trait evolution, and we plan to extend our implementation to accommodate (discrete) trait data. 

Given the potential impact of MMMs on model fit and phylogenetic reconstruction, it is important be able to examine the
realisations of such MMMs throughout evolutionary histories, which can be achieved through stochastic mapping.
Stochastic mapping of mutations or ancestral trait changes on phylogenies is a widely used method to aid evolutionary hypothesis testing \citep{Nielsen2002}.
\citet{beaulieu2013identifying} developed a stochastic mapping method for MMMs for binary traits that feature two evolutionary rate classes.
\citet{irvahn2016phylogenetic} extended this approach to allow for an arbitrary number of evolutionary rate classes, exploiting efficient sampling techniques that avoid costly matrix exponentiations \citep{irvahn2014phylogenetic}.
We will build on this work to develop stochastic mapping methods for more general MMMs that can accommodate traits that may assume an arbitrary number of states.


\section{Supplementary Material}

We have made available an online tutorial on how to construct XML files to perform phylogenetic inference using Markov-modulated models in BEAST: \url{http://beast.community/markov_modulated.html}

\section{Acknowledgments}

We are grateful to C\'ecile An\'e for kindly providing the plant plastid genome data sets.
GB acknowledges support from the Interne Fondsen KU Leuven / Internal Funds KU Leuven under grant agreement C14/18/094.
The research leading to these results has received funding from the European Research Council under the European Union's Horizon 2020 research and innovation programme (grant agreement no. 725422-ReservoirDOCS). The Artic Network receives funding from the Wellcome Trust through project 206298/Z/17/Z.
PL acknowledges support by the Research Foundation -- Flanders (`Fonds voor Wetenschappelijk Onderzoek -- Vlaanderen', G066215N, G0D5117N and G0B9317N).
MAS is partially supported through National Science Foundation grant DMS 1264153 and National Institutes of Health grants R01 AI107034 and U19 AI135995.
We gratefully acknowledge support from NVIDIA Corporation with the donation of parallel computing resources used for this research.

\section{Appendix A: simulation study}

Given the large differences in (log) marginal likelihoods obtained in favour of MMMs in the data sets we tested, for example for the \emph{psaB} and \emph{ndhD} analyses, we here assess the suitability of Bayesian model testing through (log) marginal likelihood estimation to compare MMMs to standard nucleotide substitution models.
To this end, we have simulated three data sets -- similar in complexity to the \emph{psaB} plant plastid data set -- using an updated version (1.4) of the $\pi$BUSS sequence simulation package \citep{pibuss}; we provide these XML files in the Supplementary Materials.
Each simulation was performed on a randomly drawn tree from the posterior distribution of the \emph{psaB} plant plastid data set (see section~\ref{sec:plastid}), as generated under the corresponding substitution model.
For each generated data set, we assess the performance of a range of substitution models: an independent general time-reversible model with empirical base frequencies (GTR) and estimated base frequencies (GTR+F), MMMs with two sets of base frequencies and one or two sets of symmetric rate parameters (MMM(GTR)$_{12}$ and MMM(GTR)$_{22}$), and MMMs with three sets of base frequencies and either one or three sets of symmetric rate parameters (MMM(GTR)$_{13}$ and MMM(GTR)$_{33}$).
The model fit to the data for each data set was evaluated by estimating (log) marginal likelihoods using generalized stepping-stone sampling (GSS; \citet{baele2016genealogical}), a recently introduced powerful and flexible approach to perform Bayesian model selection.
Computational demands for GSS were increased in a stepwise manner until convergence of the (log) marginal likelihoods was attained, resulting in an initial Markov chain of 20 million iterations followed by 100 power posteriors of 1 million iterations each.

\begin{figure}[!htbp]
\begin{center}
\includegraphics[scale=0.535]{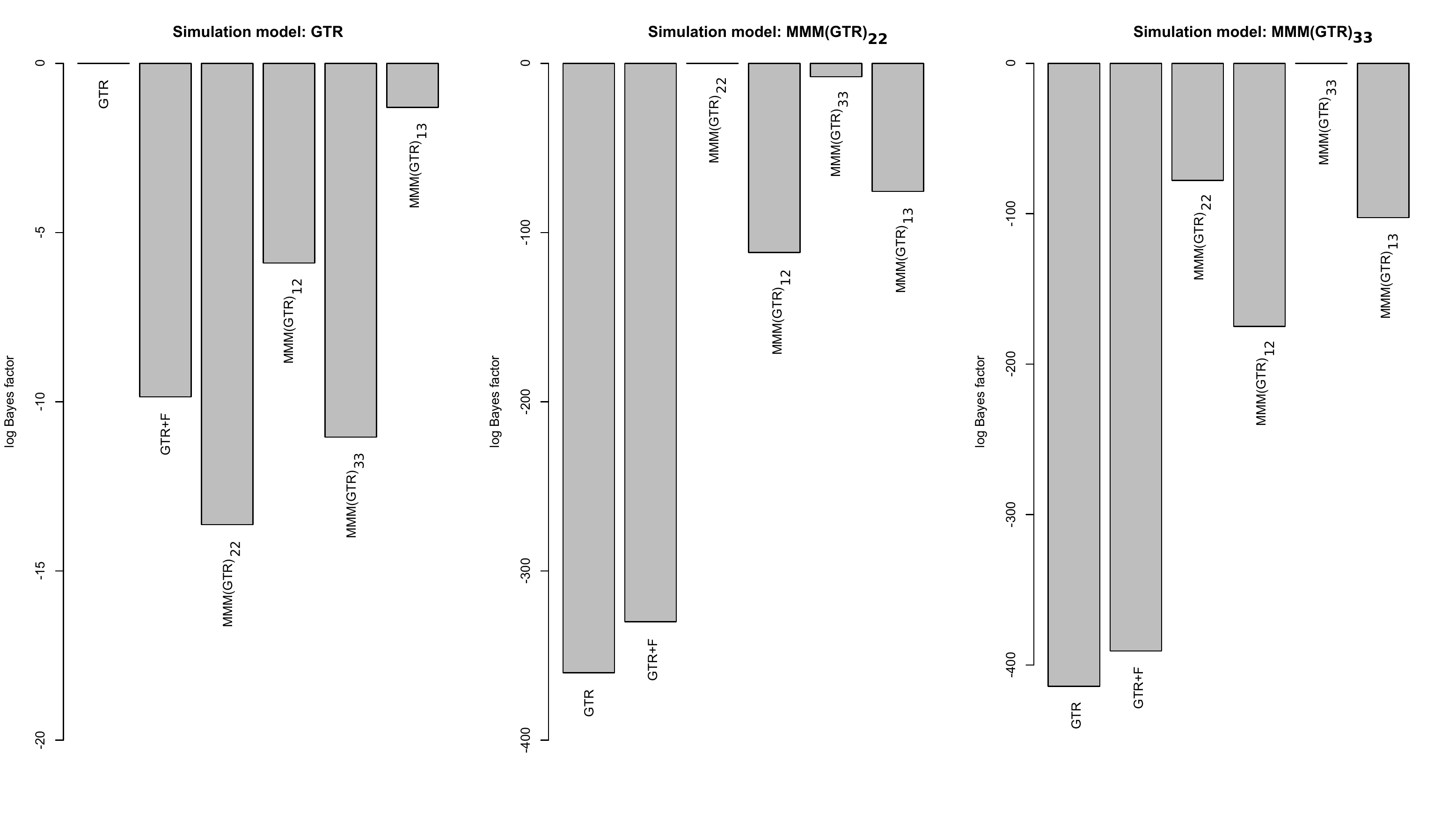}\vspace{-1cm}
\end{center}
\caption{Using a simulated data set to assess the suitability of (log) marginal likelihood estimation when performing Bayesian model selection between standard nucleotide substitution models and MMMs.
We compare the log Bayes factor of each model to that of the generative model (left: GTR; middle: MMM(GTR)$_{22}$; right: MMM(GTR)$_{33}$).
Regardless of the complexity of the model under which the data were simulated, the generative model was correctly retrieved using generalized stepping-stone sampling for (log) marginal likelihood estimation.}
\label{fig:simulation}
\end{figure}

\begin{figure}[!htbp]
\begin{center}
\includegraphics[scale=0.625]{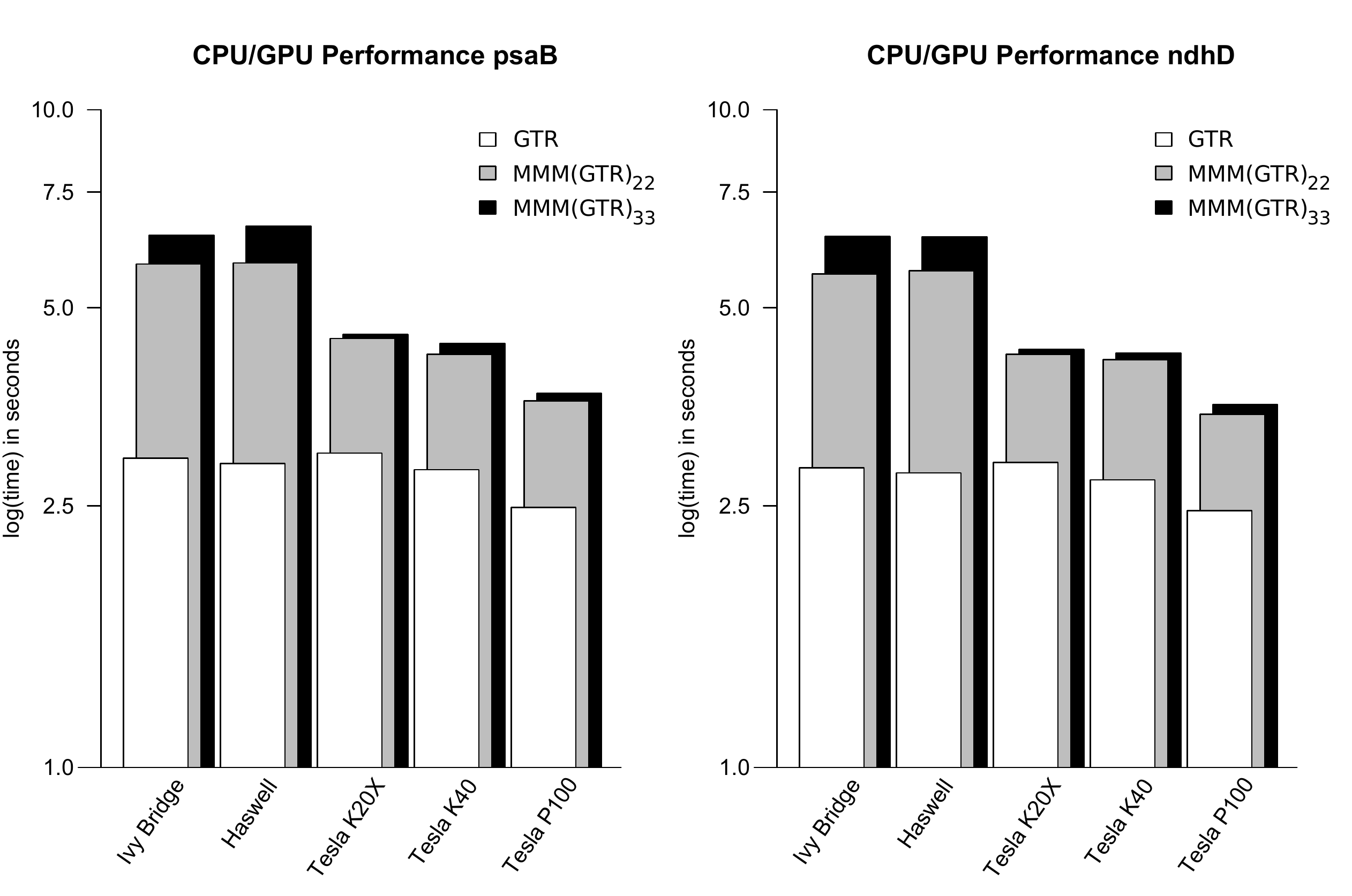}
\end{center}
\caption{Comparison of multi-core CPU and GPU hardware for the evaluation of standard nucleotide substitution models and MMMs on the \psaB and \ndhD plant plastid  data sets.
Time to completion for a short test run of 100,000 iterations using BEAST 1.10 \citep{beastX} and BEAGLE 3 \citep{beagle3} are shown for five multi-core hardware devices and are shown on a log-scale.
With the exception of the Tesla P100, all architectures deliver equal performance using a standard general time-reversible nucleotide substitution model.
CPU architectures exhibit declining performance as MMM dimensionality increases: approximately 18 and 40 times slower for MMM(GTR)$_{22}$ and MMM(GTR)$_{33}$ parameterisations, compared to a nucleotide substitution model.
GPU architectures on the other hand, limit their performance hit to a factor of approximately 4 and 5 for the MMM(GTR)$_{22}$ and MMM(GTR)$_{33}$ parameterisations, with the Tesla P100 only reporting 3 times slower estimation for those models.}
\label{fig:comp}
\end{figure}

Figure \ref{fig:simulation} summarizes the outcome of (log) marginal likelihood estimation for a range of possible substitution models under three generative models: GTR, MMM(GTR)$_{22}$ and MMM(GTR)$_{33}$.
When the nucleotide data was simulated under a GTR substitution model, (log) marginal likelihood estimation using GSS \citep{baele2016genealogical} selected the GTR model with empirical (or fixed) base frequencies as the model that best fits the data.
All other models are penalised for their abundance in parameters, illustrating that models with more parameters are not systematically preferred as the standard GTR model is a nested model within those other models.
For the data simulated under an MMM(GTR)$_{22}$, only the MMM(GTR)$_{33}$ is able to estimate similar substitution patterns, but yields a slightly lower (log) marginal likelihood, as the substitution process can be adequately captured with fewer parameters by the MMM(GTR)$_{22}$.
When the data is simulated under an MMM(GTR)$_{33}$, the generative model is retrieved with very strong (log) Bayes factor support as no simpler model is able to capture its complexity.
In conclusion, our simulations show that when a Markov-modulated process correctly captures the evolutionary patterns within a data set, (log) Bayes factor support of MMMs over standard nucleotide models will be very strong and differences in (log) marginal likelihood can amount up to hundreds of log units.

\section{Appendix B: computational performance}

Integral to computing the observed data likelihood are the finite-time transition probabilities that characterize how states transition across a branch in the phylogeny.
As discussed earlier, computation of the finite-time transition probability matrix (see Equation~\ref{eq:matrixexp}) through numerical eigendecomposition of $\boldsymbol{\Lambda}$ can be computationally demanding for MMMs owing to their high dimensionality and the lack of closed-form expressions.
While the required diagonalization of $\boldsymbol{\Lambda}$ occurs infrequently, the increased dimensionality of MMMs constitutes a computational bottleneck because of the large matrix multiplications associated with the numerical eigendecomposition.
Both diagonalization and matrix multiplications can be sped up significantly by exploiting parallelization on state-of-the-art multi-core hardware devices in order to perform matrix multiplications in parallel \citep{suchard2009}.
Recent developments in BEAST (version 1.10; \citet{beastX}) and its accompanying high-performance computational library BEAGLE (version 3; \citet{beagle3}) have brought about further improvements in dealing with such demanding use cases, particularly in the presence of multiple partitions.

We benchmark MMMs with varying dimensions against standard nucleotide substitution models, on a range of multi-core CPU and GPU hardware configurations.
In terms of raw computing power, highly parallel computing technologies such as GPUs have overtaken traditional CPUs, offering a very cost-effective option to achieve peak performance, with an increasing amount of on-board memory and ever-faster memory bandwidth.
We include two multi-core CPU configurations: a 20-core (i.e. $2 \times 10$) Intel Xeon (R) Xeon E5-2680 v2 system (Ivy Bridge microarchitecture) and a 24-core (i.e. $2 \times 12$) Intel Xeon (R) Xeon E5-2680 v3 system (Haswell microarchitecture), the former with 60 GB/s memory bandwidth and the latter with 68GB/s memory bandwidth.
We stack these CPU architectures up against three GPU devices: a Tesla K20X with 2688 CUDA cores and memory bandwidth of 250 GB/s, a Tesla K40 with 2880 CUDA cores and memory bandwidth of 288 GB/s; and a Tesla P100 with 3584 CUDA cores and memory bandwidth of 720 GB/s.

Figure~\ref{fig:comp} shows a performance comparison of these various architectures on the \emph{psaB} and \emph{ndhD} plant plastid data sets analysed in this paper.
Across the board, the Tesla P100 GPU outperforms all other hardware configurations, both for standard nucleotide substitution models and MMMs.
CPU architectures offer poor performance when evaluating MMMs and are greatly outperformed by GPU cards aimed at the scientific computing market, with graphics cards that are several years old outperforming multi-core CPUs nearly tenfold.
Comparing the performance among different GPUs shows that recent developments in the GPU market continue to offer staggering performance improvements, with the Tesla P100 GPU substantially outperforming its predecessors and drastically limiting the loss of performance associated with MMMs.
More recently developed GPUs, such as the Tesla V100, promise to deliver even further performance enhancements, but could not yet be included into this study due to lack of availability.

\bibliography{mm}

\begin{thebibliography}{53}
\providecommand{\natexlab}[1]{#1}
\providecommand{\selectlanguage}[1]{\relax}
\providecommand{\bibAnnoteFile}[1]{%
  \IfFileExists{#1}{\begin{quotation}\noindent\textsc{Key:} #1\\
  \textsc{Annotation:}\ \input{#1}\end{quotation}}{}}
\providecommand{\bibAnnote}[2]{%
  \begin{quotation}\noindent\textsc{Key:} #1\\
  \textsc{Annotation:}\ #2\end{quotation}}

\bibitem[{An\'e et~al.(2005)An\'e, Burleigh, McMahon, and Sanderson}]{ane2005}
An\'e, C., J.~G. Burleigh, M.~M. McMahon, and M.~J. Sanderson. 2005. Covarion
  structure in plastid genome evolution: A new statistical test. Mol. Biol.
  Evol. 22:914--924.
\bibAnnoteFile{ane2005}

\bibitem[{Ayres et~al.(2019)Ayres, Cummings, Baele, Darling, Lewis, Swofford,
  Huelsenbeck, Lemey, Rambaut, and Suchard}]{beagle3}
Ayres, D.~L., M.~P. Cummings, G.~Baele, A.~E. Darling, P.~O. Lewis, D.~L.
  Swofford, J.~P. Huelsenbeck, P.~Lemey, A.~Rambaut, and M.~A. Suchard. 2019.
  {BEAGLE} 3: Improved performance, scaling, and usability for a
  high-performance computing library for statistical phylogenetics. Syst. Biol.
  (in press).
\bibAnnoteFile{beagle3}

\bibitem[{Baele et~al.(2017)Baele, Lemey, Rambaut, and Suchard}]{Baele2017}
Baele, G., P.~Lemey, A.~Rambaut, and M.~A. Suchard. 2017. Adaptive {MCMC} in
  {B}ayesian phylogenetics: an application to analyzing partitioned data in
  {BEAST}. Bioinformatics 33:1798--1805.
\bibAnnoteFile{Baele2017}

\bibitem[{Baele et~al.(2016)Baele, Lemey, and Suchard}]{baele2016genealogical}
Baele, G., P.~Lemey, and M.~A. Suchard. 2016. Genealogical working
  distributions for {B}ayesian model testing with phylogenetic uncertainty.
  Syst. Biol. 65:250--264.
\bibAnnoteFile{baele2016genealogical}

\bibitem[{Beal et~al.(2002)Beal, Ghahramani, and Rasmussen}]{Beal2002}
Beal, M.~J., Z.~Ghahramani, and C.~E. Rasmussen. 2002. The infinite hidden
  {M}arkov model. Pages~577--584 \emph{in} Advances in Neural Information
  Processing Systems.
\bibAnnoteFile{Beal2002}

\bibitem[{Beaulieu et~al.(2013)Beaulieu, O'Meara, and
  Donoghue}]{beaulieu2013identifying}
Beaulieu, J.~M., B.~C. O'Meara, and M.~J. Donoghue. 2013. Identifying hidden
  rate changes in the evolution of a binary morphological character: the
  evolution of plant habit in campanulid angiosperms. Syst. Biol. 62:725--737.
\bibAnnoteFile{beaulieu2013identifying}

\bibitem[{Bielejec et~al.(2014)Bielejec, Lemey, Carvalho, Baele, Rambaut, and
  Suchard}]{pibuss}
Bielejec, F., P.~Lemey, L.~M. Carvalho, G.~Baele, A.~Rambaut, and M.~A.
  Suchard. 2014. $\pi${BUSS}: a parallel {BEAST}/{BEAGLE} utility for sequence
  simulation under complex evolutionary scenarios. BMC Bioinformatics 15:133.
\bibAnnoteFile{pibuss}

\bibitem[{Blanquart and Lartillot(2006)}]{Blanquart2006}
Blanquart, S. and N.~Lartillot. 2006. A {B}ayesian compound stochastic process
  for modeling nonstationary and nonhomogeneous sequence evolution. Mol. Biol.
  Evol. 23:2058---2071.
\bibAnnoteFile{Blanquart2006}

\bibitem[{Embley et~al.(1993)Embley, Thomas, and Williams}]{Embley1993}
Embley, T.~M., R.~H. Thomas, and R.~A.~D. Williams. 1993. Reduced thermophilic
  bias in the 16s r{DNA} sequence from {T}hermus ruber provides further support
  for a relationship between {T}hermus and {D}einococcus. System. Appl.
  Microbial. 16:25--29.
\bibAnnoteFile{Embley1993}

\bibitem[{Fan et~al.(2011)Fan, Wu, Chen, Kuo, and Lewis}]{Fan}
Fan, Y., R.~Wu, M.~H. Chen, L.~Kuo, and P.~O. Lewis. 2011. Choosing among
  partition models in {B}ayesian phylogenetics. Mol. Biol. Evol. 28:523--532.
\bibAnnoteFile{Fan}

\bibitem[{Felsenstein(1981)}]{Felsenstein1981}
Felsenstein, J. 1981. Evolutionary trees from {DNA} sequences: a maximum
  likelihood approach. J. Mol. Evol. 17:368--376.
\bibAnnoteFile{Felsenstein1981}

\bibitem[{Felsenstein(2004)}]{Felsenstein2004}
Felsenstein, J. 2004. Inferring phylogenies. Sinauer associates Sunderland.
\bibAnnoteFile{Felsenstein2004}

\bibitem[{Ferreira and Suchard(2008)}]{ferreira08}
Ferreira, M. A.~R. and M.~A. Suchard. 2008. Bayesian anaylsis of elasped times
  in continuous-time {M}arkov chains. Canadian Journal of Statistics
  26:355--368.
\bibAnnoteFile{ferreira08}

\bibitem[{Fischer and Meier-Hellstern(1993)}]{fischer1993markov}
Fischer, W. and K.~Meier-Hellstern. 1993. The markov-modulated poisson process
  (mmpp) cookbook. Performance Evaluation 18:149--171.
\bibAnnoteFile{fischer1993markov}

\bibitem[{Fitch and Markowitz(1970)}]{Fitch1970}
Fitch, W.~M. and E.~Markowitz. 1970. An improved method for determining codon
  variability in a gene and its application to the rate of fixation of
  mutations in evolution. Biochem. Genet. 4:579--593.
\bibAnnoteFile{Fitch1970}

\bibitem[{Foster(2004)}]{Foster2004}
Foster, P.~G. 2004. Modeling compositional heterogeneity. Syst. Biol.
  53:485--495.
\bibAnnoteFile{Foster2004}

\bibitem[{Galtier(2001)}]{galtier2001maximum}
Galtier, N. 2001. Maximum-likelihood phylogenetic analysis under a
  covarion-like model. Mol. Biol. Evol. 18:866--873.
\bibAnnoteFile{galtier2001maximum}

\bibitem[{Galtier and Jean-Marie(2004)}]{galtier2004markov}
Galtier, N. and A.~Jean-Marie. 2004. Markov-modulated markov chains and the
  covarion process of molecular evolution. Journal of Computational Biology
  11:727--733.
\bibAnnoteFile{galtier2004markov}

\bibitem[{Gascuel and Guindon(2007)}]{gascuel2007modelling}
Gascuel, O. and S.~Guindon. 2007. Modelling the variability of evolutionary
  processes. Reconstructing Evolution: new mathematical and computational
  advances 2:65--99.
\bibAnnoteFile{gascuel2007modelling}

\bibitem[{Guindon et~al.(2004)Guindon, Rodrigo, Dyer, and
  Huelsenbeck}]{guindon2004modeling}
Guindon, S., A.~G. Rodrigo, K.~A. Dyer, and J.~P. Huelsenbeck. 2004. Modeling
  the site-specific variation of selection patterns along lineages. Proceedings
  of the National Academy of Sciences of the United States of America
  101:12957--12962.
\bibAnnoteFile{guindon2004modeling}

\bibitem[{Hasegawa et~al.(1985)Hasegawa, Kishino, and Yano}]{HKY}
Hasegawa, M., H.~Kishino, and T.~Yano. 1985. Dating of the human-ape splitting
  by a molecular clock of mitochondrial {DNA}. J. Mol. Evol. 22:160--174.
\bibAnnoteFile{HKY}

\bibitem[{Huelsenbeck(2002)}]{Huelsenbeck2002}
Huelsenbeck, J.~P. 2002. Testing a covariotide model of {DNA} substitution.
  Molecular Biology and Evolution 19:698--707.
\bibAnnoteFile{Huelsenbeck2002}

\bibitem[{Irvahn(2016)}]{irvahn2016phylogenetic}
Irvahn, J. 2016. Phylogenetic Stochastic Mapping. Ph.D. thesis University of
  Washington, Department of Statistics.
\bibAnnoteFile{irvahn2016phylogenetic}

\bibitem[{Irvahn and Minin(2014)}]{irvahn2014phylogenetic}
Irvahn, J. and V.~N. Minin. 2014. Phylogenetic stochastic mapping without
  matrix exponentiation. Journal of Computational Biology 21:676--690.
\bibAnnoteFile{irvahn2014phylogenetic}

\bibitem[{Jukes and Cantor(1969)}]{Jukes1969}
Jukes, T.~H. and C.~R. Cantor. 1969. {Evolution of Protein Molecules}.
  \emph{in} Evolution of Protein Molecules (H.~N. Munro, ed.). Academy Press.
\bibAnnoteFile{Jukes1969}

\bibitem[{Lanave et~al.(1984)Lanave, Preparata, Saccone, and
  Serio}]{Lanave1984}
Lanave, C., G.~Preparata, C.~Saccone, and G.~Serio. 1984. A new method for
  calculating evolutionary substitution rates. Journal of Molecular Evolution
  20:86--93.
\bibAnnoteFile{Lanave1984}

\bibitem[{Lockhart et~al.(1998)Lockhart, Steel, Barbrook, Huson, Charleston,
  and Howe}]{Lockhart1998}
Lockhart, P.~J., M.~A. Steel, A.~C. Barbrook, D.~H. Huson, M.~A. Charleston,
  and C.~J. Howe. 1998. A covariotide model explains apparent phylogenetic
  structure of oxygenic photosynthetic lineages. Mol. Biol. Evol.
  15:1183--1188.
\bibAnnoteFile{Lockhart1998}

\bibitem[{Lopez et~al.(2002)Lopez, Casane, and Philippe}]{Lopez2002}
Lopez, P., D.~Casane, and H.~Philippe. 2002. Heterotachy, an important process
  of protein evolution. Mol. Biol. Evol. 19:1--7.
\bibAnnoteFile{Lopez2002}

\bibitem[{Mooers and Holmes(2000)}]{Mooers2000}
Mooers, A.~O. and E.~C. Holmes. 2000. The evolution of base composition and
  phylogenetic inference. Trends Ecol. Evol. 15:365--369.
\bibAnnoteFile{Mooers2000}

\bibitem[{Murray(1992)}]{murray1991}
Murray, R. G.~E. 1992. The family {D}einococcaceae. Pages~3732---3744 \emph{in}
  The Prokaryotes: A Handbook on the Biology of Bacteria: Ecophysiology,
  Isolation, Identification, Applications (A.~Balows, H.~G. Tru\"per,
  M.~Dworkin, W.~Harder, and K.-H. Schleifer, eds.) vol.~4. Springer-Verlag,
  New York.
\bibAnnoteFile{murray1991}

\bibitem[{Nielsen(2002)}]{Nielsen2002}
Nielsen, R. 2002. Mapping mutations on phylogenies. Syst. Biol. 51:729--739.
\bibAnnoteFile{Nielsen2002}

\bibitem[{Pagel and Meade(2004)}]{Pagel2004}
Pagel, M. and A.~Meade. 2004. A phylogenetic mixture model for detecting
  pattern-heterogeneity in gene sequence or character-state data. Syst. Biol.
  53:571--581.
\bibAnnoteFile{Pagel2004}

\bibitem[{Pan and Chen(1999)}]{pan1999complexity}
Pan, V.~Y. and Z.~Q. Chen. 1999. The complexity of the matrix eigenproblem.
  Pages~507--516 \emph{in} Proceedings of the Thirty-first Annual ACM Symposium
  on Theory of Computing STOC '99 ACM, New York, NY, USA.
\bibAnnoteFile{pan1999complexity}

\bibitem[{Rambaut et~al.(2018)Rambaut, Drummond, Xie, Baele, and
  Suchard}]{rambaut2018posterior}
Rambaut, A., A.~Drummond, D.~Xie, G.~Baele, and M.~Suchard. 2018. Posterior
  summarisation in {B}ayesian phylogenetics using {Tracer} 1.7. Systematic
  Biology 67:901--904.
\bibAnnoteFile{rambaut2018posterior}

\bibitem[{Rambaut et~al.(2008)Rambaut, Pybus, Nelson, Viboud, Taubenberger, and
  Holmes}]{rambaut2008}
Rambaut, A., O.~Pybus, M.~Nelson, C.~Viboud, J.~Taubenberger, and E.~Holmes.
  2008. The genomic and epidemiological dynamics of human influenza {A} virus.
  Nature 453:615--619.
\bibAnnoteFile{rambaut2008}

\bibitem[{Shapiro et~al.(2006)Shapiro, Rambaut, and Drummond}]{Shapiro2006}
Shapiro, B., A.~Rambaut, and A.~J. Drummond. 2006. Choosing appropriate
  substitution models for the phylogenetic analysis of protein-coding
  sequences. Mol. Biol. Evol. 23:7--9.
\bibAnnoteFile{Shapiro2006}

\bibitem[{Silverman(1986)}]{Silverman}
Silverman, B.~W. 1986. The kernel method for univariate data. Pages~34--72
  \emph{in} Density Estimation for Statistics and Data Analysis. Chapman \&
  Hall/CRC., London.
\bibAnnoteFile{Silverman}

\bibitem[{Singer and Ames(1970)}]{singer1970}
Singer, C.~E. and B.~N. Ames. 1970. Sunlight ultraviolet and bacterial dna base
  ratios. Science 170:822--826.
\bibAnnoteFile{singer1970}

\bibitem[{Stern and Elwalid(1991)}]{stern1991analysis}
Stern, T.~E. and A.~I. Elwalid. 1991. Analysis of separable markov-modulated
  rate models for information-handling systems. Advances in Applied Probability
  Pages~105--139.
\bibAnnoteFile{stern1991analysis}

\bibitem[{Suchard et~al.(2018)Suchard, Lemey, Baele, Ayres, Drummond, and
  Rambaut}]{beastX}
Suchard, M.~A., P.~Lemey, G.~Baele, D.~L. Ayres, A.~J. Drummond, and
  A.~Rambaut. 2018. Bayesian phylogenetic and phylodynamic data integration
  using {BEAST} 1.10. Virus Evol. 4:vey016.
\bibAnnoteFile{beastX}

\bibitem[{Suchard and Rambaut(2009)}]{suchard2009}
Suchard, M.~A. and A.~Rambaut. 2009. Many-core algorithms for statistical
  phylogenetics. Bioinformatics 25:1370--1376.
\bibAnnoteFile{suchard2009}

\bibitem[{Tavar\'{e}(1986)}]{Tavare1986}
Tavar\'{e}, S. 1986. Some probabilistic and statistical problems in the
  analysis of {DNA} sequences. Pages~57--86 \emph{in} Some mathematical
  questions in biology: {DNA} sequence analysis. (M.~S. Waterman, ed.).
  American Mathematical Society., Providence (RI).
\bibAnnoteFile{Tavare1986}

\bibitem[{Tolver(2016)}]{Tolver2016}
Tolver, A. 2016. An introduction to {M}arkov chains.
\bibAnnoteFile{Tolver2016}

\bibitem[{Tuffley and Steel(1998)}]{tuffley1998modeling}
Tuffley, C. and M.~Steel. 1998. Modeling the covarion hypothesis of nucleotide
  substitution. Mathematical Biosciences 147:63--91.
\bibAnnoteFile{tuffley1998modeling}

\bibitem[{Wang et~al.(2007)Wang, Spencer, Susko, and Roger}]{Wang2007}
Wang, H.-C., M.~Spencer, E.~Susko, and A.~J. Roger. 2007. Testing for
  covarion-like evolution in protein sequences. Mol. Biol. Evol. 24:294--305.
\bibAnnoteFile{Wang2007}

\bibitem[{Wang et~al.(2009)Wang, Susko, and Roger}]{Wang2009}
Wang, H.-C., E.~Susko, and A.~J. Roger. 2009. {PROCOV:} maximum likelihood
  estimation of protein phylogeny under covarion models and site-specific
  covarion pattern analysis. BMC Evolutionary Biology 9:225.
\bibAnnoteFile{Wang2009}

\bibitem[{Whelan(2008)}]{Whelan2008}
Whelan, S. 2008. Spatial and temporal heterogeneity in nucleotide sequence
  evolution. Mol. Biol. Evol. 25:1683--1694.
\bibAnnoteFile{Whelan2008}

\bibitem[{Worobey et~al.(2014)Worobey, Han, and Rambaut}]{Worobey2014}
Worobey, M., G.~Z. Han, and A.~Rambaut. 2014. A synchronized global sweep of
  the internal genes of modern avian influenza virus. Nature 508:254--257.
\bibAnnoteFile{Worobey2014}

\bibitem[{Wu et~al.(2013)Wu, Suchard, and Drummond}]{Wu2013}
Wu, C.-H., M.~A. Suchard, and A.~J. Drummond. 2013. Bayesian selection of
  nucleotide substitution models and their site assignments. Mol. Biol. Evol.
  30:669--688.
\bibAnnoteFile{Wu2013}

\bibitem[{Yang(1994)}]{yang1994maximum}
Yang, Z. 1994. Maximum likelihood phylogenetic estimation from {DNA} sequences
  with variable rates over sites: approximate methods. J. Mol. Evol.
  39:306--314.
\bibAnnoteFile{yang1994maximum}

\bibitem[{Yang(1996)}]{Yang1996}
Yang, Z. 1996. Among-site rate variation and its impact on phylogenetic
  analyses. Trends Ecol. Evol. 11:367--372.
\bibAnnoteFile{Yang1996}

\bibitem[{Yule(1924)}]{Yule1924}
Yule, G.~U. 1924. A mathematical theory of evolution based on the conclusions
  of {Dr}. {J}.{C}. willis. F.R.S. Philos. Trans. R. Soc. Lond. B Biol. Sci.
  213:21--87.
\bibAnnoteFile{Yule1924}

\bibitem[{Zhou et~al.(2010)Zhou, Brinkmann, Rodrigue, Lartillot, and
  Philippe}]{Zhou2010}
Zhou, Y., H.~Brinkmann, N.~Rodrigue, N.~Lartillot, and H.~Philippe. 2010. A
  {D}irichlet process covarion mixture model and its assessments using
  posterior predictive discrepancy tests. Mol. Biol. Evol. 27:371--384.
\bibAnnoteFile{Zhou2010}

\end{thebibliography}

\end{document}